\documentclass[11pt,a4paper,pdftex]{article}
\pdfoutput=1
\usepackage{jcappub}

\newcommand{\be}{\begin{eqnarray}}
\newcommand{\ee}{\end{eqnarray}}
\newcommand{\rar}{\rightarrow}

\title{Probing the space-time geometry around black hole candidates with the resonance models for high-frequency QPOs and comparison with the continuum-fitting method}

\author{Cosimo Bambi}

\affiliation{Arnold Sommerfeld Center for Theoretical Physics,\\
Ludwig-Maximilians-Universit\"at M\"unchen,\\ D-80333 Munich, Germany}

\emailAdd{Cosimo.Bambi@physik.uni-muenchen.de}

\abstract{Astrophysical black hole candidates are thought to be the Kerr black hole 
predicted by General Relativity. However, in order to confirm the Kerr-nature of these 
objects, we need to probe the geometry of the space-time around them and check that
observations are consistent with the predictions of the Kerr metric. That can be 
achieved, for instance, by studying the properties of the electromagnetic radiation 
emitted by the gas in the accretion disk. The high-frequency quasi-periodic oscillations 
observed in the X-ray flux of some stellar-mass black hole candidates might do the 
job. As the frequencies of these oscillations depend only very weakly on the observed 
X-ray flux, it is thought they are mainly determined by the metric of the space-time. In 
this paper, I consider the resonance models proposed by Abramowicz and Kluzniak 
and I extend previous results to the case of non-Kerr space-times. The emerging 
picture is more complicated than the one around a Kerr black hole and there is a larger 
number of possible combinations between different modes. I then compare the 
bounds inferred from the twin peak high-frequency quasi-periodic oscillations observed
in three micro-quasars (GRO~J1655-40, XTE~J1550-564, and GRS~1915+105)
with the measurements from the continuum-fitting method of the same objects. For 
Kerr black holes, the two approaches do not provide consistent results. In a non-Kerr
geometry, this conflict may be solved if the observed quasi-periodic oscillations are 
produced by the resonance $\nu_\theta : \nu_r = 3:1$, where $\nu_\theta$ and $\nu_r$ 
are the two epicyclic frequencies. It is at least worth mentioning that the deformation 
from the Kerr solution required by observations would be consistent with the one 
suggested in another recent work discussing the possibility that steady jets are 
powered by the spin of these compact objects.}

\begin{document}

\maketitle


\section{Introduction}

In 4-dimensional General Relativity, uncharged black holes (BHs) are described
by the Kerr solution, which is completely specified by two quantities: the mass,
$M$, and the spin parameter, $a_* = J/M^2$, where $J$ is the spin angular
momentum, of the compact object\footnote{Throughout the paper, I use units in 
which $G_{\rm N} = c = 1$.}~\cite{nh1,nh2,nh3}. A fundamental limit for a Kerr 
BH is the bound $|a_*| \le 1$, which is the condition for the existence of the event 
horizon. For $|a_*| > 1$, there is no horizon and the Kerr metric describes the 
gravitational field around a naked singularity, which is forbidden  by the Weak 
Cosmic Censorship Conjecture~\cite{wccc}.

It is widely believed that the final product of the gravitational collapse is a Kerr 
BH, as follows from a body of analytical and numerical studies~\cite{wccc,bh1,bh2,bh3,bh4} 
(but see~\cite{ns1,ns2,loop,loop2} and references therein). Astronomers
have also discovered several good astrophysical candidates~\cite{nara}. 
In particular, we have about twenty stellar-mass BH candidates in X-ray binary 
systems. For the time being, we can only get a robust measurement of the mass 
of these objects. That is possible by studying the orbital motion of the stellar 
companion. The latter is relatively far from the compact object and its trajectory 
is well described by Newtonian mechanics. In this way, we can estimate the mass 
of the compact object without any assumption about its nature. As these masses
turn out to exceed 3~$M_\odot$ (the maximum mass for a neutron or quark 
star assuming no phase transition to exotic forms of matter at densities below 
the nuclear one~\cite{max1,max2}), these objects are thought to be Kerr BHs, as 
they cannot be explained otherwise without introducing new physics.

In order to confirm the Kerr-nature of the known astrophysical BH candidates, we
need to probe the geometry of the space-time around them and check if it is
consistent with the predictions of General Relativity~\cite{review}. That can be 
achieved in a number of way. In the case of stellar-mass BH candidates, we may
get information on the geometry of the space-time around these objects by 
studying the thermal spectrum of their accretion disk during the high-soft 
state~\cite{cfm1,cfm2}, by the analysis of relativistic lines~\cite{fabian,iron}, and 
possibly even by measuring the power of their jets~\cite{j1,j2}. If the stellar 
companion is a radio pulsar, its orbital motion can be accurately tracked and we 
can get information about the nature of the BH candidate~\cite{wex}. 
Ground-based gravitational wave detectors are supposed to be able to
observe BH quasi-normal modes (QNMs) in a near future and, since the ones
of a Kerr BH depend only on $M$ and $J$, the possible detection of at least 
three modes can test the Kerr-nature of the source~\cite{qnm}. Other 
approaches can instead be used only to probe the geometry of the space-time 
around super-massive BH candidates in galactic nuclei~\cite{gw1,gw2,gw3,will,sha1,sha2,sha3,sha4,rad1,rad2}.

Another promising tool to test the nature of stellar-mass BH candidates is 
represented by the so-called quasi-periodic oscillations (QPOs). The X-ray power 
density spectra of some low-mass X-ray binaries show some peaks; that is, 
there are QPOs in the X-ray flux\footnote{QPOs have been observed even 
in the spectra of super-massive BH candidates, 
see e.g. Ref.~\cite{gie}.}~\cite{mcc}. These features can be observed in systems 
with both BH candidates and neutron stars and the frequencies of these oscillations
are in the range 0.1-10$^3$~Hz. High-frequency QPOs in BH candidates
($\sim$40-450~Hz) are particularly interesting, as they depend only very weakly
on the observed X-ray flux and for this reason it is thought they are determined
by the metric of the space-time rather than by the properties of the accretion flow.
If that is correct, high-frequency QPOs can be used to probe the geometry around 
stellar-mass BH candidates. For the time being, however, the exact physical mechanism
responsible for the production of the high-frequency QPOs is not known and several
different scenarios have been proposed, including hot-spot models~\cite{hsm1,hsm2},
diskoseismology models~\cite{dm1,dm2,dm3,dm4}, and resonance 
models~\cite{rm1,rm2,ka05,rm3}. In these models, the frequencies of the QPOs are
directly related to the characteristic orbital frequencies of test-particles.

The possibility of using QPOs to test the strong gravitational field around stellar-mass 
BH candidates has been already studied~\cite{qpo1,qpo2,qpo3}. 
In~\cite{qpo1,qpo2}, the authors consider a very specific case, the braneworld 
BH metric proposed in~\cite{brane}, in which the geometry of the space-time is
fairly similar to the one of an electrically charged BH of General Relativity. Let us 
notice, however, that this solution may be strongly constrained by current Solar 
System experiments (but only if the Birkhoff's Theorem holds). In Ref.~\cite{qpo3}, 
the authors study QPOs in a quasi-Kerr metric, valid only for small spin parameters 
and small deviations from the Kerr solution. In this paper, I study the phenomenon 
in a quite generic stationary and axisymmetric non-Kerr background and I find a 
much more complicated picture. In particular, the possibility of the existence of
vertically unstable orbits, absent in the Kerr background, allows for a larger
number of possible combinations between different modes. I then consider the
scenario of resonance models proposed by Abramowicz and Kluzniak~\cite{rm1,rm2,ka05,rm3},
mainly motivated by the observed 3:2 double peak high-frequency QPOs
in micro-quasars, and I compare theoretical predictions with observational data.
Lastly, I compare these results with the measurements inferred from the continuum-fitting
method of the same objects. In the Kerr case, this approach is equivalent to
the comparison of the spin measurements from high-frequency QPOs
and continuum-fitting method and the question of consistency of the two
techniques has been discussed in Ref.~\cite{tkss}. While high-frequency QPOs
and continuum-fitting method do not seem to provide consistent results for Kerr BHs, 
the conflict may be solved if the space-time around BH candidates deviates from 
the Kerr solution.

The content of the paper is as follows. In Sec.~\ref{s-2}, I discuss the characteristic
orbital frequencies of test-particles in the Johannsen-Psaltis (JP) metric~\cite{jp},
which is a stationary and axisymmetric space-time with very generic deviations
from the Kerr background. In Sec.~\ref{s-3}, I apply the results of the previous section
to the resonance models. In Sec.~\ref{s-4}, I consider the observational data of
the three stellar-mass BH candidates with observed twin peak high-frequency QPOs and
with a measurement of the mass. These objects are GRO~J1655-40, XTE~J1550-564, 
and GRS~1915+105. Fortunately, for them we have also measurements of
the thermal spectrum of their disk during the high-soft state and I can thus compare
the allowed region in the plane spin parameter-deformation parameter inferred
from the QPOs with the one determined from the continuum-fitting method. Discussion 
and conclusions are in Sec.~\ref{s-5}.

\section{Characteristic orbital frequencies \label{s-2}}

Let us now focus on equatorial orbits, as the accretion disk around BH candidates
is normally expected to lie on the equatorial plane of the system. Circular orbits 
of test-particles are characterized by three frequencies: the Keplerian frequency 
$\nu_{\rm K}$ (which is the inverse of the orbital period), the radial epicyclic 
frequency $\nu_r$ (the frequency of radial oscillations around the mean orbit), and 
the vertical epicyclic frequency $\nu_\theta$ (the frequency of vertical oscillations 
around the mean orbit). These three frequencies depend on the geometry of the 
space-time and on the radius of the orbit. While they are defined as the characteristic 
frequencies of the orbital motion for a free particle, there is a direct relation between 
these frequencies and the ones of the oscillation modes of the fluid accretion flow.

\begin{figure}
\par
\begin{center}
\includegraphics[type=pdf,ext=.pdf,read=.pdf,width=7.5cm]{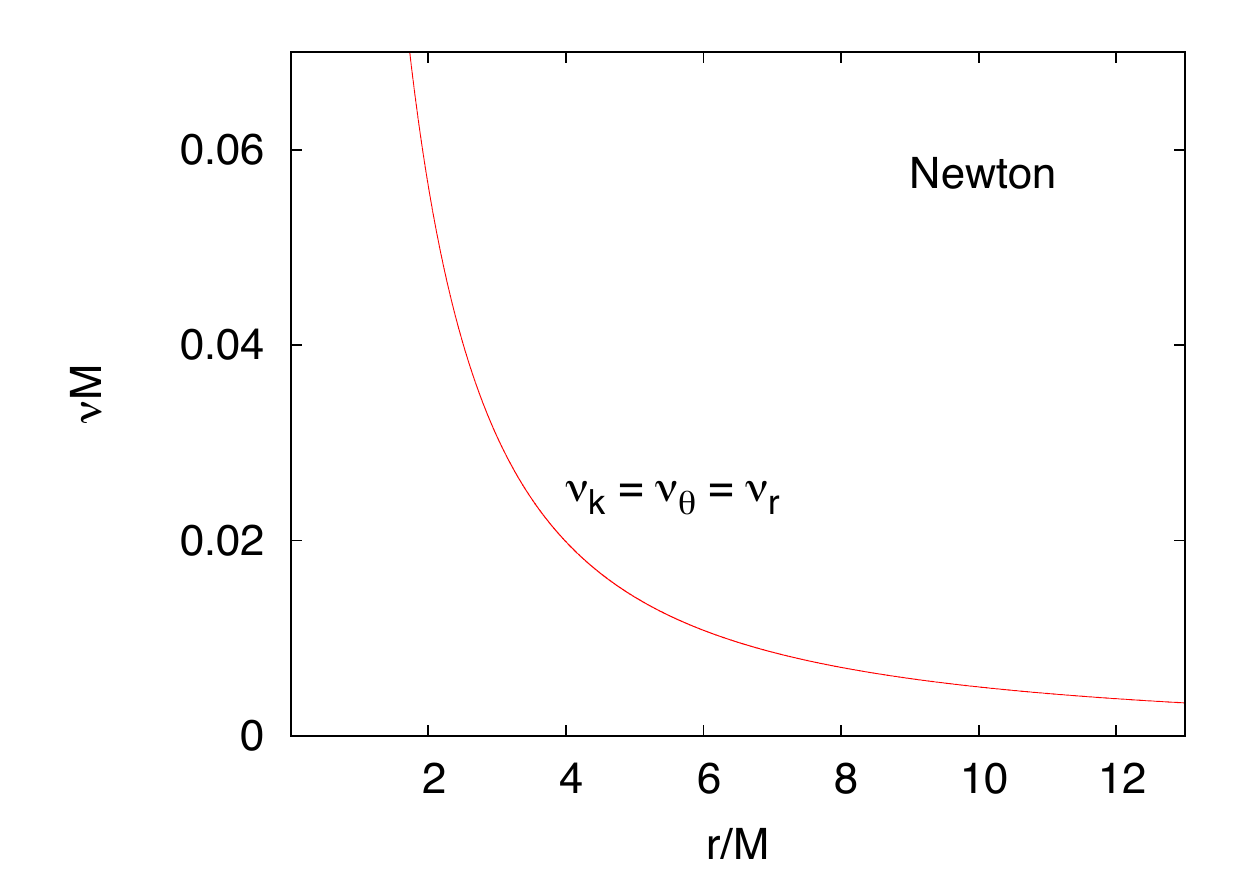}
\includegraphics[type=pdf,ext=.pdf,read=.pdf,width=7.5cm]{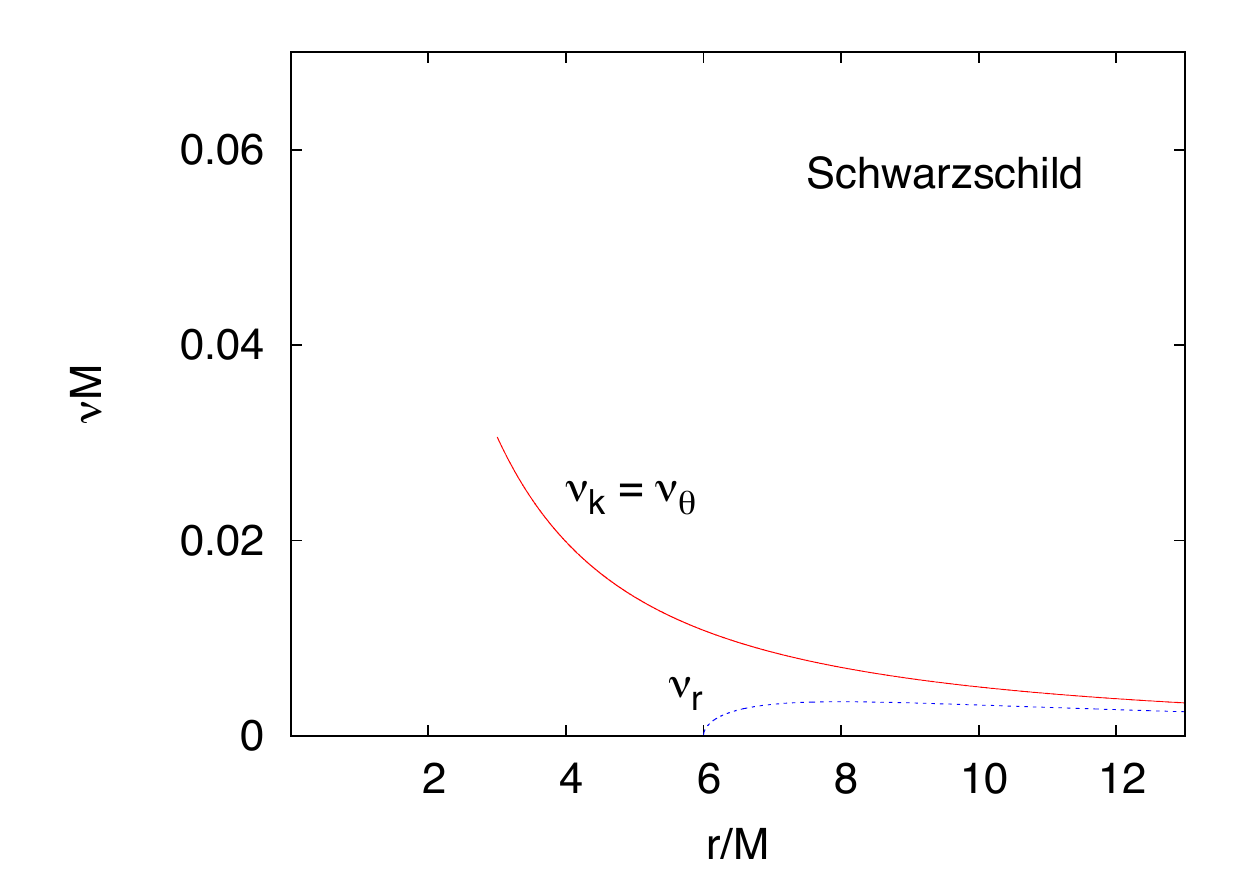} \\
\includegraphics[type=pdf,ext=.pdf,read=.pdf,width=7.5cm]{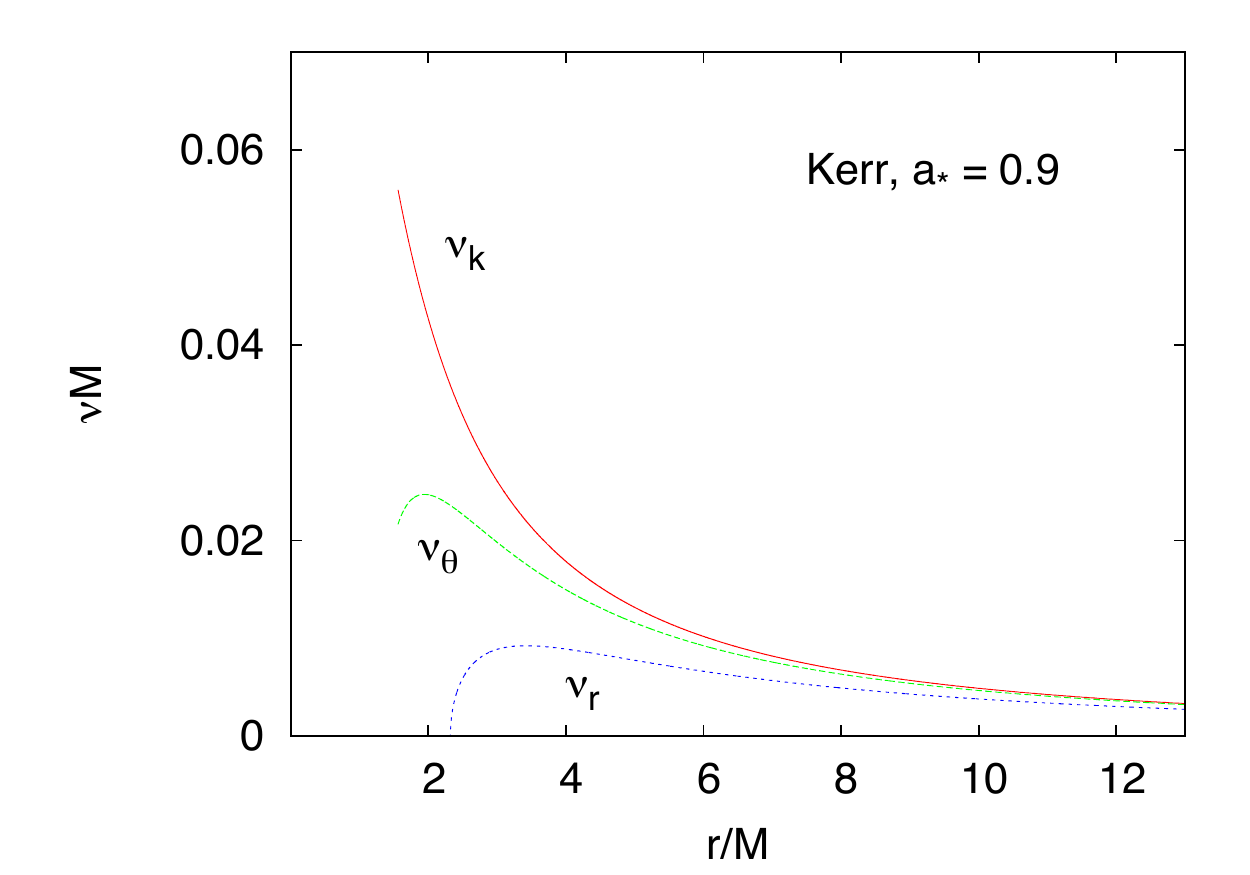}
\includegraphics[type=pdf,ext=.pdf,read=.pdf,width=7.5cm]{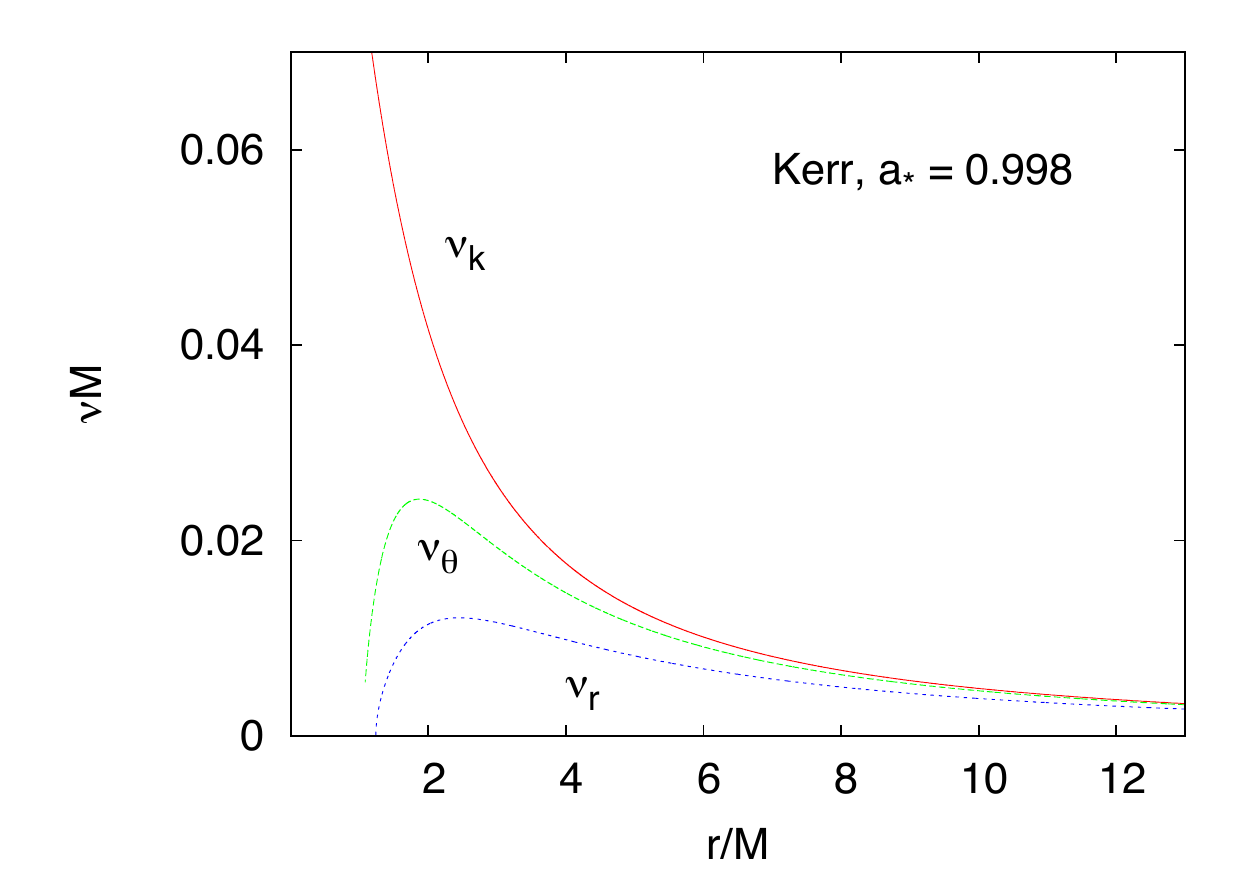} 
\end{center}
\par
\vspace{-5mm} 
\caption{Profiles of the Keplerian frequency $\nu_{\rm K}$, the radial epicyclic 
frequency $\nu_r$, and the vertical epicyclic frequency $\nu_\theta$  in Newtonian 
gravity with potential $- M / r$ (top left panel), in the Schwarzschild space-time 
(top right panel), and in the Kerr space-time with spin parameter $a_* = 0.9$ 
(bottom left panel) and $a_* = 0.998$ (bottom right panel). See text for details.}
\label{f-1}
\end{figure}

In Newtonian gravity with potential $V = - M/r$, the three characteristic 
frequencies has the same value:
\be
\nu_{\rm K} = \nu_r = \nu_\theta = \frac{M}{r^{3/2}} \, .
\ee
They are plotted as a function of the radial coordinate in the top left panel of 
Fig.~\ref{f-1}.

In General Relativity, a key-ingredient is the existence of the innermost stable 
circular orbit (ISCO). In the Schwarzschild metric, the ISCO radius is $r_{\rm ISCO} = 6 M$. 
In the Kerr background, the ISCO radius decreases (increases) as the spin 
parameter $a_*$ increases for corotating (counterrotating) orbits. For a maximally 
rotating Kerr BH ($a_*=1$), the ISCO radius is respectively $r_{\rm ISCO} = M$ 
for the corotating case and $r_{\rm ISCO} = 9 M$ for the counterrotating one. Circular 
orbits with radii smaller than $r_{\rm ISCO}$ are radially unstable. That means the
radial epicyclic frequency $\nu_r$ reaches a maximum at some radius $r_{\rm max} > 
r_{\rm ISCO}$ and then vanishes at the ISCO. Keplerian and vertical epicyclic frequencies
are instead defined up to the innermost circular orbit (also called photon orbit).
Circular orbits with radius smaller than the one of the photon orbit do not exist. In the 
Kerr metric, $\nu_\theta > \nu_r$, while, for corotating orbits, $\nu_{\rm K} \ge \nu_\theta$.
$\nu_{\rm K}$, $\nu_r$, and $\nu_\theta$ in the Schwarzschild space-time are shown 
in the top right panel of Fig.~\ref{f-1}. The Kerr cases with spin parameter $a_* = 0.9$ 
and $0.998$ and for corotating orbits are shown respectively in the bottom left and 
bottom right panels of Fig.~\ref{f-1}.

In the case of a generic non-Kerr background, the picture is more complicated.
The most important difference is that the ISCO radius may be determined by the
stability of the orbit along the vertical (instead of the radial) direction~\cite{cfm1,bb2}. 
In some metrics, there is also the possibility of the existence of one or more regions 
of stable circular orbits inside the ISCO and therefore with smaller radii. These
regions of stable orbits are separated by a gap from the ``traditional'' ISCO; the
details depend on the specific metric, but usually they do not play any physical role, 
as they can accept only particles with energy higher than the one at the ISCO. In what 
follows, I will consider the JP metric~\cite{jp}, as it seems to include all the relevant 
features of a generic space-time deviating from the Kerr solution. 
While several authors have proposed specific alternatives to Kerr BHs for the
astrophysical BH candidates (like Q-stars, gravastars, or non-Kerr BHs solutions 
of specific and theoretically motivated extensions of General Relativity), here the 
choice of the JP metric with a single deformation parameter can be motivated as
follows. The aim of this work is not to test a specific theoretical model and
to determine/constrain its parameters, but to investigate in a model-independent 
way possible deviations from the Kerr geometry of the space-time around
astrophysical BHs. The JP metric is used to perform a null-experiment: it is like 
the Kerr metric with a deformation parameter measuring possible deviations from 
the Kerr solution and the spirit is to determine this deformation parameter and 
check it is zero. Current and near-future data are indeed not so good to map 
the space-time around a BH candidate and a single deformation parameter is 
used to figure out if the gravitational force is stronger or weaker than the one 
around a Kerr BH with the same mass and spin. Actually, the typical situation 
is even worse and with one measurement we can only infer one parameter: if 
we assume the Kerr background, we find the spin $a_*$ (but we find a wrong 
value if the object is not a Kerr BH), if we have also a deformation parameter,
we constrain some combination of $a_*$ and of the deformation parameter.
With two independent measurements, we can determine both the spin and 
the deformation parameter. The choice of the JP metric should thus be
understood with this spirit and it makes sense if it is used to check if the 
geometry of the space-time is described by the Kerr solution, not to investigate 
the actual nature of the compact object or the exact deviations from the Kerr 
background.

In Boyer-Lindquist coordinates, the JP metric is given by the line element
\be\label{eq-jp}
ds^2 &=& - \left(1 - \frac{2 M r}{\Sigma}\right) (1 + h) \, dt^2 + \nonumber\\
&& + \frac{\Sigma (1 + h)}{\Delta + a^2 h \sin^2\theta } \, dr^2 
+ \Sigma \, d\theta^2 - \frac{4 a M r \sin^2\theta}{\Sigma} (1 + h) \, dt \, d\phi +
\nonumber\\ && + \left[\sin^2\theta \left(r^2 + a^2 
+ \frac{2 a^2 M r \sin^2\theta}{\Sigma} \right)
+ \frac{a^2 (\Sigma + 2 M r) \sin^4\theta}{\Sigma} h \right] d\phi^2 \, ,
\ee
where $a = a_* M$, $\Sigma = r^2 + a^2 \cos^2\theta$,
$\Delta = r^2 - 2 M r + a^2$, and
\be
h = \sum_{k = 0}^{\infty} \left(\epsilon_{2k}
+ \frac{M r}{\Sigma} \epsilon_{2k+1} \right)
\left(\frac{M^2}{\Sigma}\right)^k \, .
\ee
The JP metric has an infinite number of deformation parameters $\epsilon_i$ and
the Kerr solution is recovered when all the deformation parameters are set to
zero. However, in order to reproduce the correct Newtonian limit, we have to
impose $\epsilon_0 = \epsilon_1 = 0$, while $\epsilon_2$ is strongly constrained 
by Solar System experiments~\cite{jp}. In this paper, I consider a simple case
with a sole deformation parameter $\epsilon_3$ and $\epsilon_i = 0$ for $i \neq 3$.
A different choice of the deformation parameter, like $\epsilon_4$ or $\epsilon_5$ 
instead of $\epsilon_3$, would not change the qualitative features of our non-Kerr
metric, as well as our results and conclusions~\cite{agn}.

In a generic stationary, axisymmetric, and asymptotically flat space-time, the 
characteristic orbital frequencies can be computed numerically as follows. 
The line element of the space-time can be written in the canonical form
\be
ds^2 = g_{tt} dt^2 + g_{rr}dr^2 + g_{\theta\theta} d\theta^2 
+ 2g_{t\phi}dt d\phi + g_{\phi\phi}d\phi^2 \, ,
\ee
where the metric components are independent of the $t$ and $\phi$ coordinates, 
which implies the existence of two constants of motion: the conserved specific 
energy at infinity, $E$, and the conserved $z$-component of the specific angular 
momentum at infinity, $L_z$. This fact allows to write the $t$- and $\phi$-component 
of the 4-velocity of a test-particle as 
\be
u^t = \frac{E g_{\phi\phi} + L_z g_{t\phi}}{
g_{t\phi}^2 - g_{tt} g_{\phi\phi}} \, , \qquad 
u^\phi = - \frac{E g_{t\phi} + L_z g_{tt}}{
g_{t\phi}^2 - g_{tt} g_{\phi\phi}} \, .
\ee
From the conservation of the rest-mass, $g_{\mu\nu}u^\mu u^\nu = -1$,
we can write
\be
g_{rr}\dot{r}^2 + g_{\theta\theta}\dot{\theta}^2
= V_{\rm eff}(r,\theta,E,L_z) \, ,
\ee
where $\dot{r} = u^r = dr/d\lambda$, $\dot{\theta} = u^\theta = d\theta/d\lambda$,
$\lambda$ is an affine parameter, and the effective potential $V_{\rm eff}$ is given by
\be
V_{\rm eff} = \frac{E^2 g_{\phi\phi} + 2 E L_z g_{t\phi} + L^2_z 
g_{tt}}{g_{t\phi}^2 - g_{tt} g_{\phi\phi}} - 1  \, .
\ee
Circular orbits on the equatorial plane are located at the zeros and the turning 
points of the effective potential: $\dot{r} = \dot{\theta} = 0$, which implies 
$V_{\rm eff} = 0$, and $\ddot{r} = \ddot{\theta} = 0$, requiring respectively 
$\partial_r V_{\rm eff} = 0$ and $\partial_\theta V_{\rm eff} = 0$. From these 
conditions, one can obtain the Keplerian angular velocity $\Omega_{\rm K}$, $E$, 
and $L_z$ of the test-particle:
\be
\Omega_{\rm K} &=& \frac{d\phi}{dt} = 
\frac{- \partial_r g_{t\phi} 
\pm \sqrt{\left(\partial_r g_{t\phi}\right)^2 
- \left(\partial_r g_{tt}\right) \left(\partial_r 
g_{\phi\phi}\right)}}{\partial_r g_{\phi\phi}} \, , \\
E &=& - \frac{g_{tt} + g_{t\phi}\Omega}{
\sqrt{-g_{tt} - 2g_{t\phi}\Omega - g_{\phi\phi}\Omega^2}} \, , \\
L_z &=& \frac{g_{t\phi} + g_{\phi\phi}\Omega}{
\sqrt{-g_{tt} - 2g_{t\phi}\Omega - g_{\phi\phi}\Omega^2}} \, ,
\ee
where in $\Omega_{\rm K}$ the sign $+$ is for corotating orbits 
and the sign $-$ for counterrotating 
ones. The Keplerian frequency is simply $\nu_{\rm K} = \Omega_{\rm K}/2\pi$.
The orbits are stable under small perturbations if $\partial_r^2 V_{\rm eff} \le 0$ 
and $\partial_\theta^2 V_{\rm eff} \le 0$.

\begin{figure}
\par
\begin{center}
\includegraphics[type=pdf,ext=.pdf,read=.pdf,width=7.5cm]{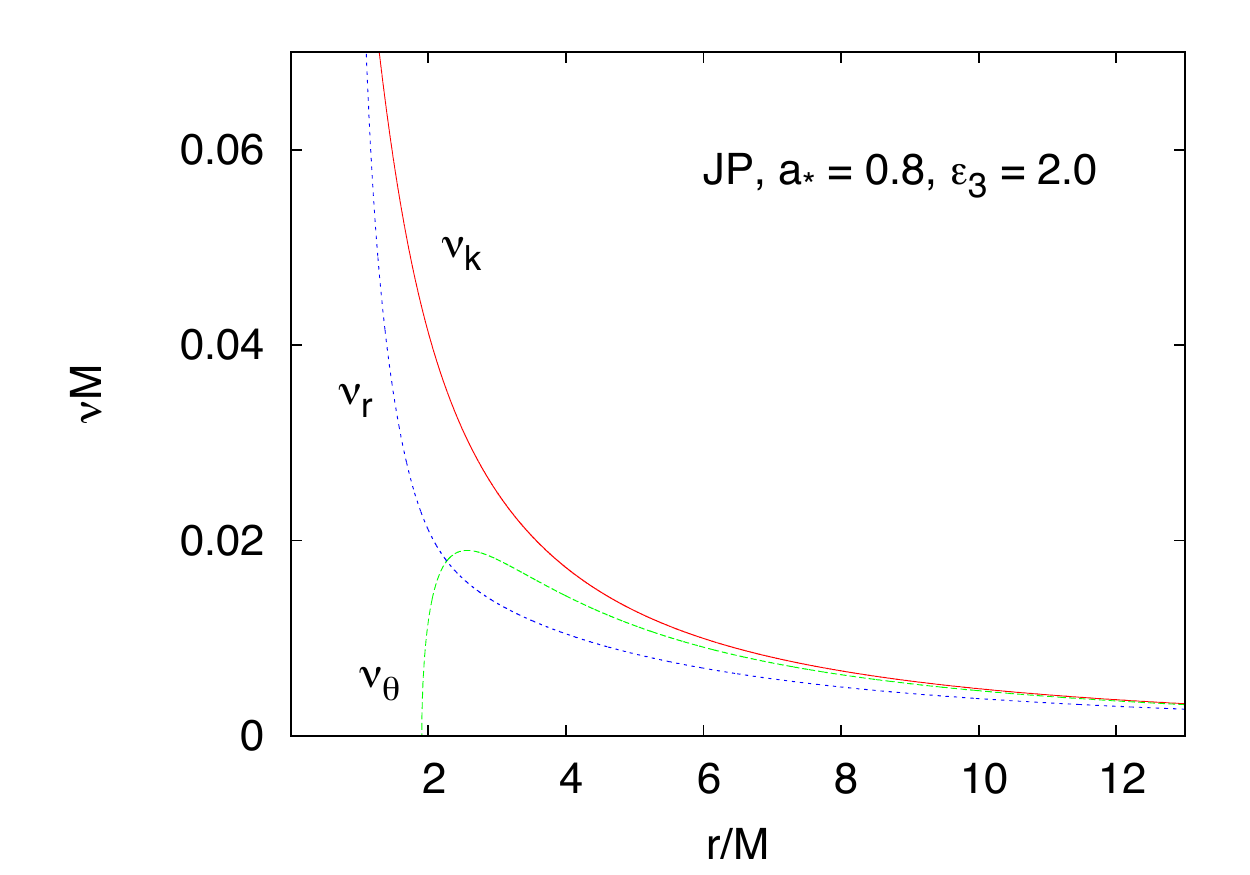}
\includegraphics[type=pdf,ext=.pdf,read=.pdf,width=7.5cm]{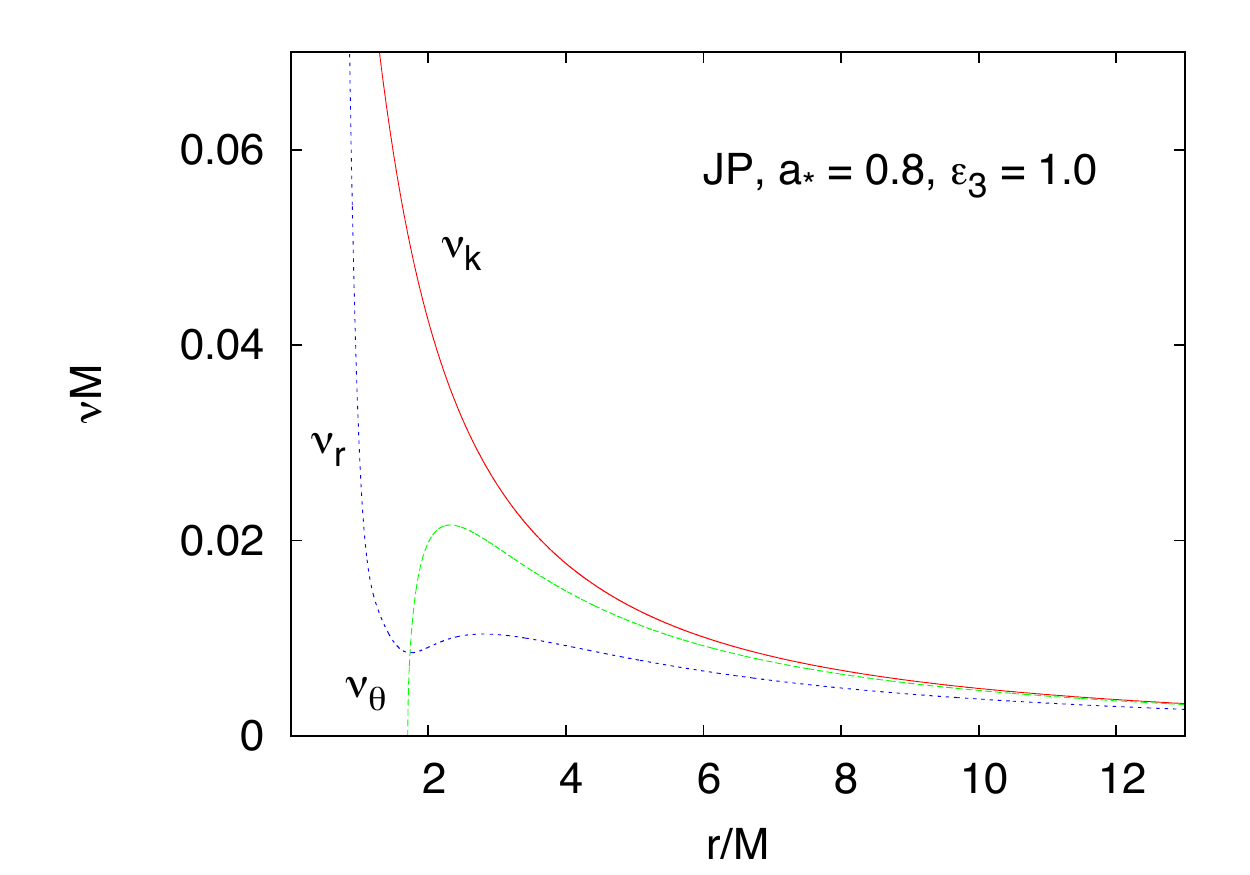} \\
\includegraphics[type=pdf,ext=.pdf,read=.pdf,width=7.5cm]{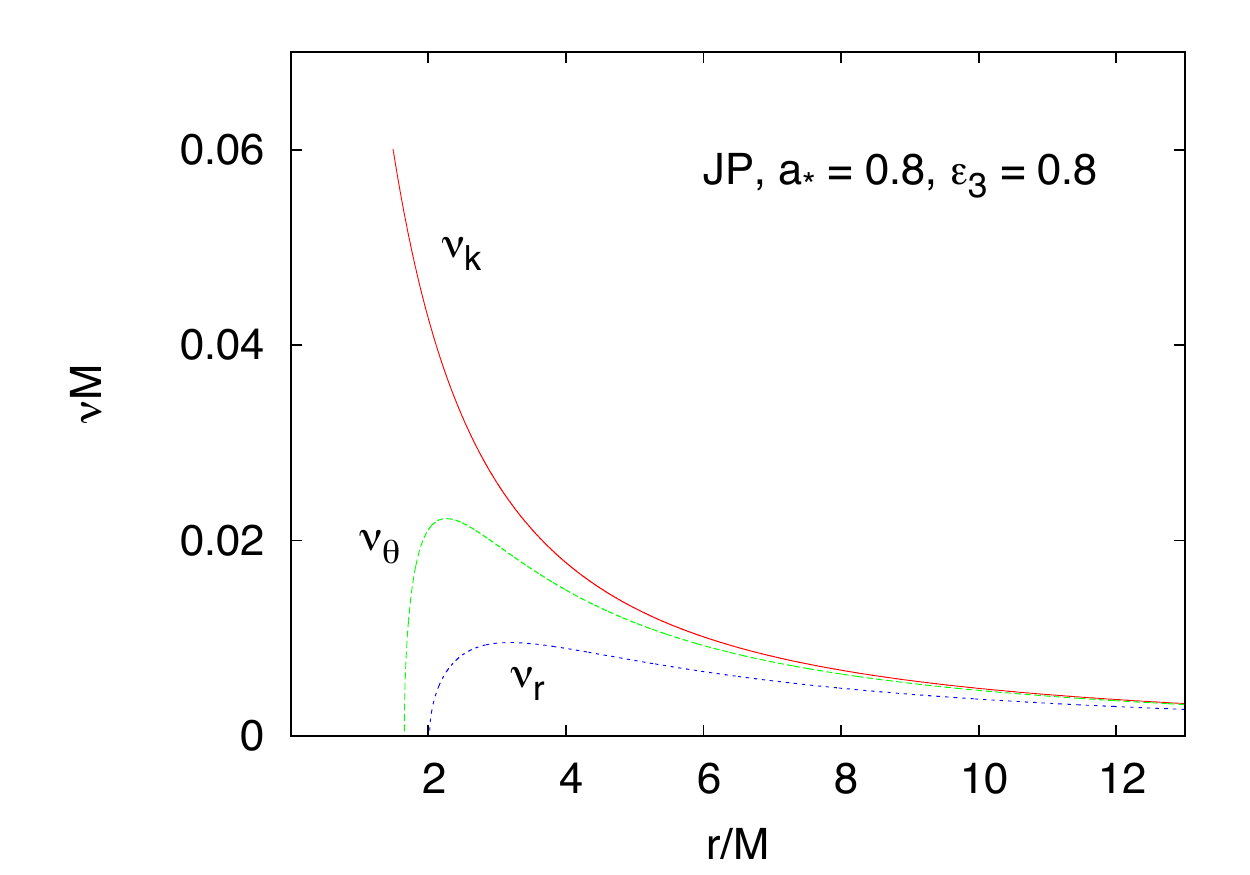}
\includegraphics[type=pdf,ext=.pdf,read=.pdf,width=7.5cm]{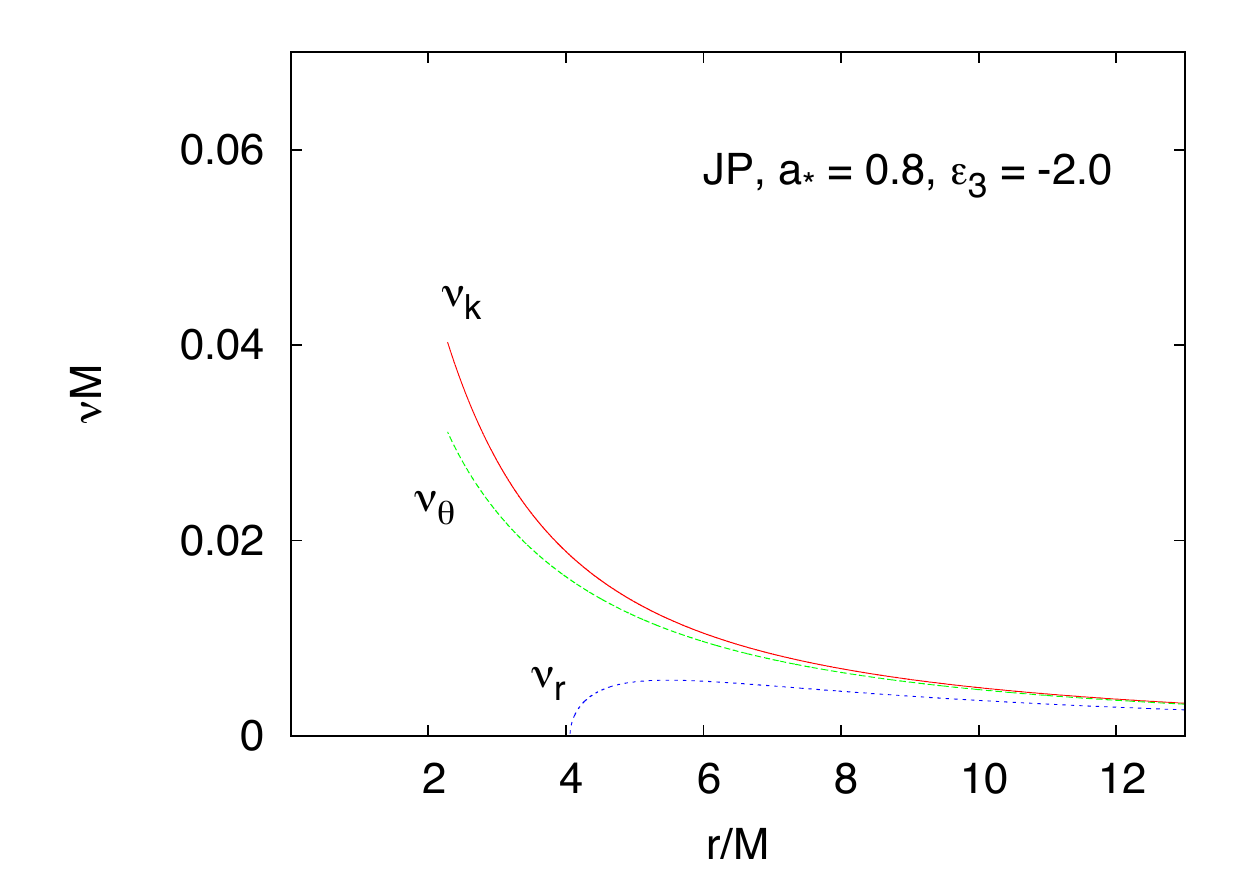} 
\end{center}
\par
\vspace{-5mm} 
\caption{Profiles of the Keplerian frequency $\nu_{\rm K}$, the radial epicyclic 
frequency $\nu_r$, and the vertical epicyclic frequency $\nu_\theta$ in the JP 
space-time with spin parameter $a_* = 0.8$ and deformation parameter 
$\epsilon_3 = 2.0$ (top left panel), $\epsilon_3 = 1.0$ (top right panel), 
$\epsilon_3 = 0.8$ (bottom left panel), and $\epsilon_3 = -2.0$ (bottom right 
panel). See text for details.}
\label{f-2}
\end{figure}

The radial and vertical epicyclic frequencies can be quickly computed by considering
small perturbations around circular equatorial orbits, respectively along the radial and
vertical direction. If $\delta_r$ and $\delta_\theta$ are the small displacements around
the mean orbit (i.e. $r = r_0 + \delta_r$ and $\theta = \pi/2 + \delta_\theta$), 
we find they are governed by the following differential equations
\be\label{eq-o1}
\frac{d^2 \delta_r}{dt^2} + \Omega_r^2 \delta_r &=& 0 \, , \\
\frac{d^2 \delta_\theta}{dt^2} + \Omega_\theta^2 \delta_\theta &=& 0 \, ,
\label{eq-o2}
\ee
where~\cite{ka05}
\be\label{eq-or}
\Omega^2_r &=& - \frac{1}{2 g_{rr} (u^t)^2} 
\frac{\partial^2 V_{\rm eff}}{\partial r^2} \, , \\
\Omega^2_\theta &=& - \frac{1}{2 g_{\theta\theta} (u^t)^2} 
\frac{\partial^2 V_{\rm eff}}{\partial \theta^2} \, .
\label{eq-ot}
\ee
The radial epicyclic frequency is thus $\nu_r = \Omega_r/2\pi$ and the vertical one 
is $\nu_\theta = \Omega_\theta/2\pi$. The radial profile of the three characteristic 
orbital frequencies in the JP background with spin parameter $a_* = 0.8$ and
four different values of the deformation parameter $\epsilon_3$ is shown in
Fig.~\ref{f-2}.

\section{Resonance models \label{s-3}}

In four stellar-mass BH candidates, we observe two high-frequency QPOs. It is 
remarkable that in all the four cases the ratio of the two frequencies is 3:2,
suggesting a strong correlation between them. Possible theoretical models
should thus be able to explain this feature. The resonance models proposed 
by Abramowicz and Kluzniak~\cite{rm1,rm2,ka05,rm3} seem to be quite appealing
from this point of view. In Eqs.~(\ref{eq-o1}) and (\ref{eq-o2}), the radial and 
the vertical modes are decoupled. However, it is natural to expect that in a 
more realistic description there are non-linear effects coupling the two epicyclic 
modes. In this case, the equations can be written as 
\be\label{eq-o3}
\frac{d^2 \delta_r}{dt^2} + \Omega_r^2 \delta_r &=& 
\Omega_r^2 F_r \left( \delta_r, \delta_\theta, \frac{d \delta_r}{dt}, 
\frac{d \delta_\theta}{dt}\right) \, , \\
\frac{d^2 \delta_\theta}{dt^2} + \Omega_\theta^2 \delta_\theta &=& 
\Omega_\theta^2 F_\theta \left( \delta_r, \delta_\theta, \frac{d \delta_r}{dt}, 
\frac{d \delta_\theta}{dt}\right) \, ,
\ee
where $F_r$ and $F_\theta$ are some functions that depend on the specific 
properties of the accretion flow. If we knew the details of the physical
mechanisms of the accretion process, we could write the explicit form of these
two functions and solve the system. Unfortunately, that is not the case.
The strategy is therefore to guess possible properties and consequences
of these equations and see if they can be fitted in a plausible physical scenario.

\subsection{Parametric resonances \label{ss-me}}

A simple but interesting scenario is to imagine that vertical oscillations are 
governed by the Mathieu equation~\cite{rm2,rm3}:
\be\label{eq-o5}
\frac{d^2 \delta_\theta}{dt^2} + \Omega_\theta^2 \delta_\theta &=& 
- \Omega_\theta^2 h \cos(\Omega_r t) \delta_\theta \, ,
\ee
which corresponds to the case with $F_r = 0$ and $F_\theta = - \delta_r \delta_\theta$:
the solution of Eq.~(\ref{eq-o3}) is simply $\delta_r = h \cos(\Omega_r t)$ and so
we obtain Eq.~(\ref{eq-o5}). The Mathieu equation describes indeed a parametric
resonance with
\be
\frac{\Omega_r}{\Omega_\theta} = \frac{2}{n} \, , \qquad n = 1, 2, 3, . . .
\ee
The resonance is stronger for smaller values of $n$. In the Kerr background, $\nu_\theta 
> \nu_r$, and therefore the resonance $n = 3$ can naturally explain the observed
3:2 ratio if the upper frequency $\nu_{\rm U}$ is associated with $\nu_\theta$ 
and the lower frequency $\nu_{\rm L}$ with $\nu_r$. In the JP space-time with 
$\epsilon_3 > 0$, $\nu_\theta > \nu_r$ may not be true (depending on the value
of the spin parameter), as the ISCO may be determined by the orbital stability
along the vertical direction. In this case, the resonances $n = 1$ and 2 may be
exited and indeed be stronger than the resonance $n = 3$. For $n =1$, the observed
3:2 ratio might be interpreted as $\nu_{\rm U} = \nu_r + \nu_\theta$ and 
$\nu_{\rm L} = \nu_r$. For $n = 2$, as $\nu_{\rm U} = 3 \nu_r = 3 \nu_\theta$ 
and $\nu_{\rm L} = 2 \nu_r = 2 \nu_\theta$.

\subsection{Forced resonances}

The equation for vertical oscillations may also include a forced resonance, in which
the force frequency is equal to the one of the radial oscillations~\cite{rm3}:
\be
\frac{d^2 \delta_\theta}{dt^2} + \Omega_\theta^2 \delta_\theta 
+ [{\rm non \; linear \; terms \; in} \: \delta_\theta] &=& 
h(r) \cos(\Omega_r t) \, .
\ee
The non-linear terms allow resonant solutions for $\delta_\theta$,
with frequencies like
\be
\Omega_- &=& \Omega_\theta - \Omega_r \, , \\
\Omega_+ &=& \Omega_\theta + \Omega_r \, .
\ee
The observed 3:2 ratio may be explained with $\nu_\theta : \nu_r = 3:1$
($\nu_{\rm U} = \nu_\theta$ and $\nu_{\rm L} = \nu_-$) or with $\nu_\theta : \nu_r = 2:1$
($\nu_{\rm U} = \nu_+$ and $\nu_{\rm L} = \nu_\theta$). In non-Kerr backgrounds
with vertically unstable ISCO, we have also the possibility that $\nu_r > \nu_\theta$
and therefore, at least in principle, resonances like $\nu_\theta : \nu_r = 1:2$ and
$\nu_\theta : \nu_r = 1:3$ may exist.

\subsection{Keplerian resonances}

The possibility of a coupling between Keplerian and radial epicyclic frequencies might
exist, even if it seems to be less theoretically motivated than the one in which 
the coupling is between the two epicyclic oscillations. The simplest combinations are:
$\nu_{\rm K} : \nu_r = 3:2$ ($\nu_{\rm U} = \nu_{\rm K}$ and $\nu_{\rm L} = \nu_r$),
$\nu_{\rm K} : \nu_r = 3:1$ ($\nu_{\rm U} = \nu_{\rm K}$ and $\nu_{\rm L} = 2 \nu_r$),
and
$\nu_{\rm K} : \nu_r = 2:1$ ($\nu_{\rm U} = 3 \nu_r$ and $\nu_{\rm L} = \nu_{\rm K}$).
For non-Kerr backgrounds, there are no new phenomena with respect to the standard
Kerr framework, as $\nu_{\rm K} > \nu_r$ is still true.

\section{The observed twin peak high-frequency QPOs in BH candidates \label{s-4}}

In four stellar-mass BH candidates we observe two high-frequency QPOs in the X-ray
flux and the ratio between the upper and the lower frequency turns out to be 
$\nu_{\rm U} : \nu_{\rm L} = 3 : 2$. For three of these objects, we have also a
dynamical measurement of the mass, which allows us to use the observed twin peak
high-frequency QPOs to test their nature (supposing we have the exact model for the
production of QPOs). These three objects are listed in Tab.~\ref{tab}, with their mass
$M$ and the observed upper and lower high-frequency QPOs $\nu_{\rm U}$ and 
$\nu_{\rm L}$. Regardless of the specific resonance model and the microphysics
responsible for it, the resonance paradigm requires that the upper and the lower
frequencies have the form
\be
\nu_{\rm U} &=& m_1 \nu_r + m_2 \nu_\theta \, , \\
\nu_{\rm L} &=& n_1 \nu_r + n_2 \nu_\theta \, . 
\ee
where $m_1$, $m_2$, $n_1$, and $n_2$ are integer (and likely as small as 
possible) numbers. For Keplerian resonance models, $\nu_{\rm U}$ and $\nu_{\rm L}$
have clearly the form $\nu_{\rm U} = m_1 \nu_{\rm K} + m_2 \nu_r$ and $\nu_{\rm L} 
= n_1 \nu_{\rm K} + n_2 \nu_r$. Assuming the JP background with deformation 
parameter $\epsilon_3$, we can compare the theoretical predictions with the observed
$\nu_{\rm U}$ and $\nu_{\rm L}$ for any BH candidate and define an allowed 
region on the plane spin parameter-deformation parameter for any specific set
of $\{ m_1, m_2, n_1, n_2 \}$.

\begin{table}
\begin{center}
\begin{tabular}{c c c c c c c c c}
\hline \\
BH binary & \hspace{.5cm} & $M/M_\odot$ & \hspace{.5cm} & 
$\nu_{\rm U}$/Hz & \hspace{.5cm} & $\nu_{\rm L}$/Hz & \hspace{.5cm} & References \\ \\
\hline
GRO~J1655-40 & & $6.30 \pm 0.27$ & & $450 \pm 3$ & & $300 \pm 5$ & & \cite{stro01} \\
XTE~J1550-564 & & $9.1 \pm 0.6$ & & $276 \pm 3$ & & $184 \pm 5$ & & \cite{rem02} \\ 
GRS~1915+105 & & $14.0 \pm 4.4$ & & $168 \pm 3$ & & $113 \pm 5$ & & \cite{mcc} \\
\hline
\end{tabular}
\end{center}
\caption{Stellar-mass BH candidates in binary systems with a measurement of the 
mass and two observed high-frequency QPOs.}
\label{tab}
\end{table}

\subsection{GRO~J1655-40}

\begin{figure}
\par
\begin{center}
\includegraphics[type=pdf,ext=.pdf,read=.pdf,width=7.5cm]{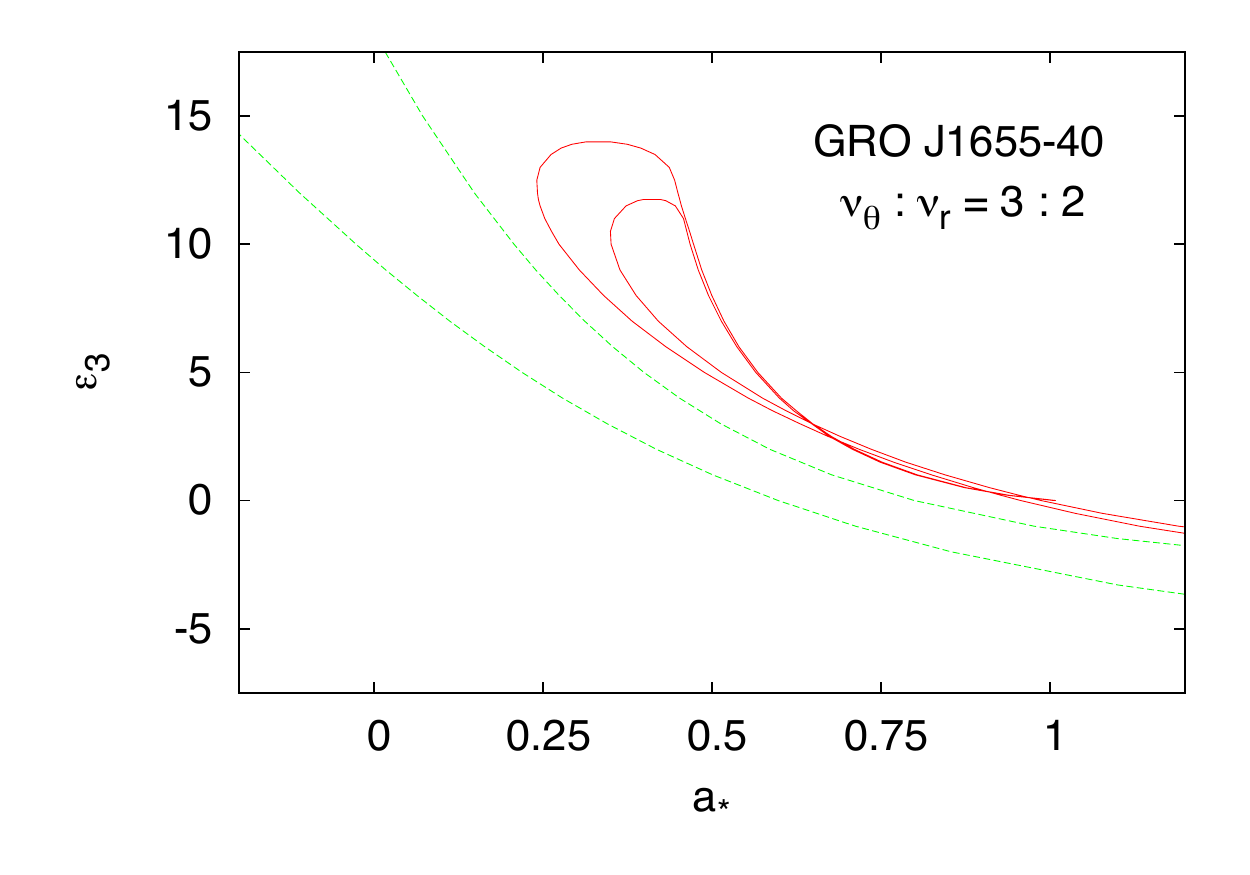}
\includegraphics[type=pdf,ext=.pdf,read=.pdf,width=7.5cm]{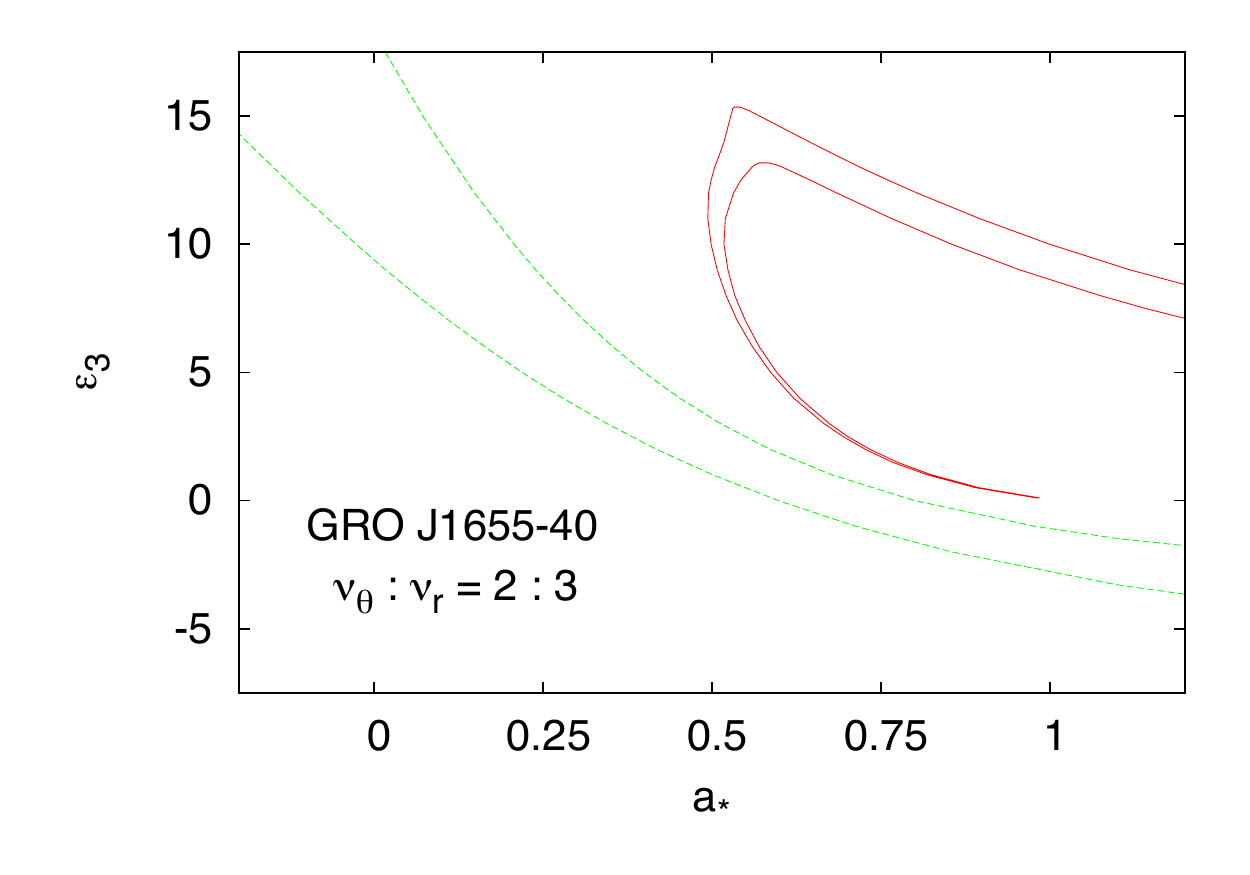} \\
\includegraphics[type=pdf,ext=.pdf,read=.pdf,width=7.5cm]{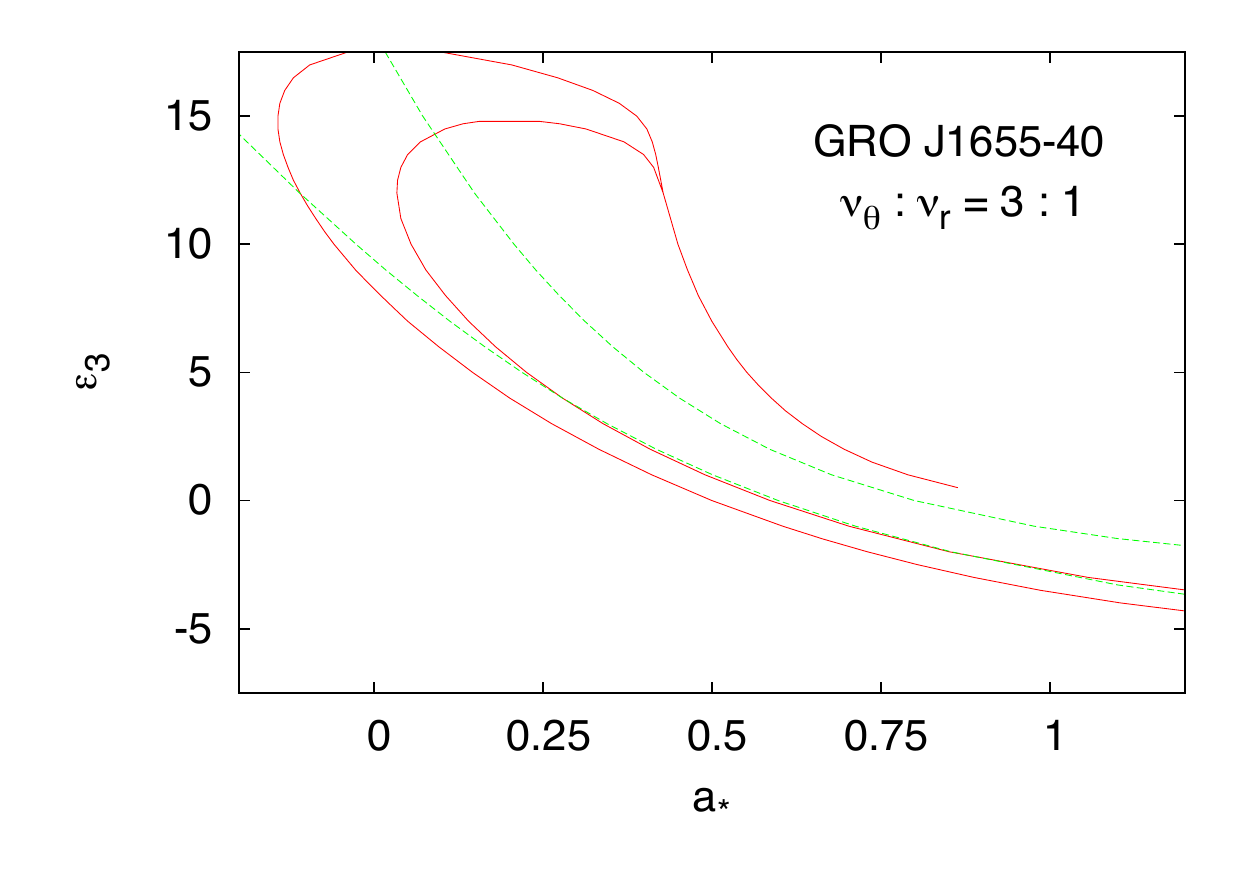} 
\includegraphics[type=pdf,ext=.pdf,read=.pdf,width=7.5cm]{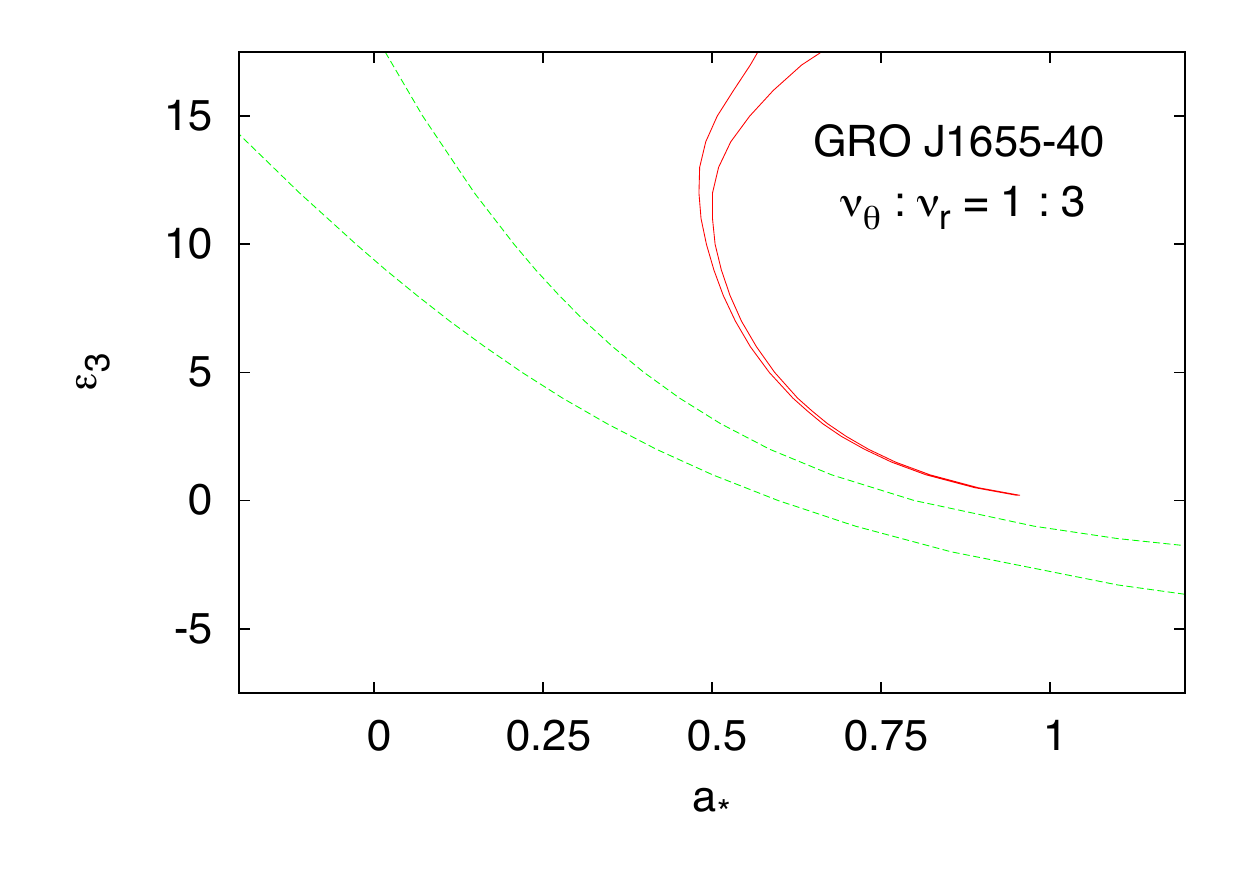} \\
\includegraphics[type=pdf,ext=.pdf,read=.pdf,width=7.5cm]{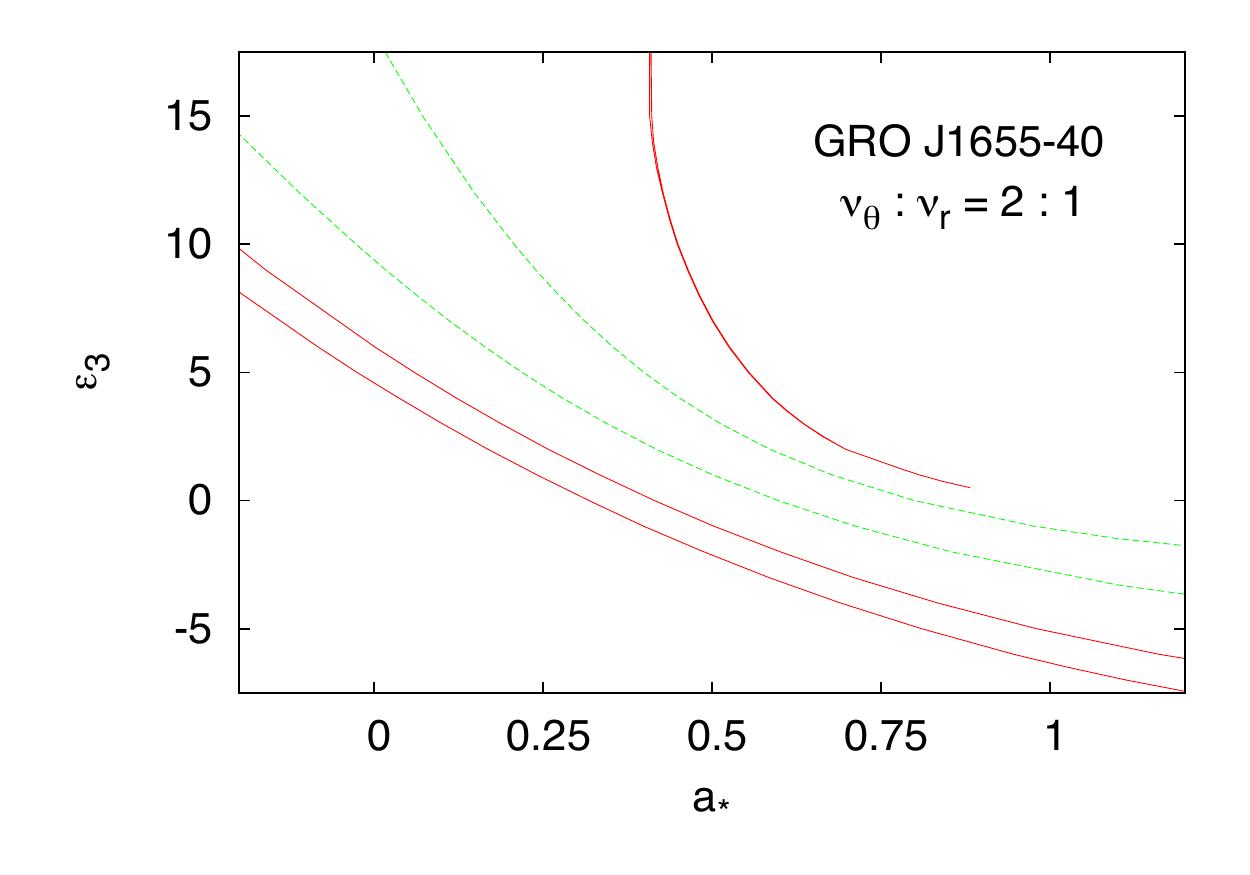} 
\includegraphics[type=pdf,ext=.pdf,read=.pdf,width=7.5cm]{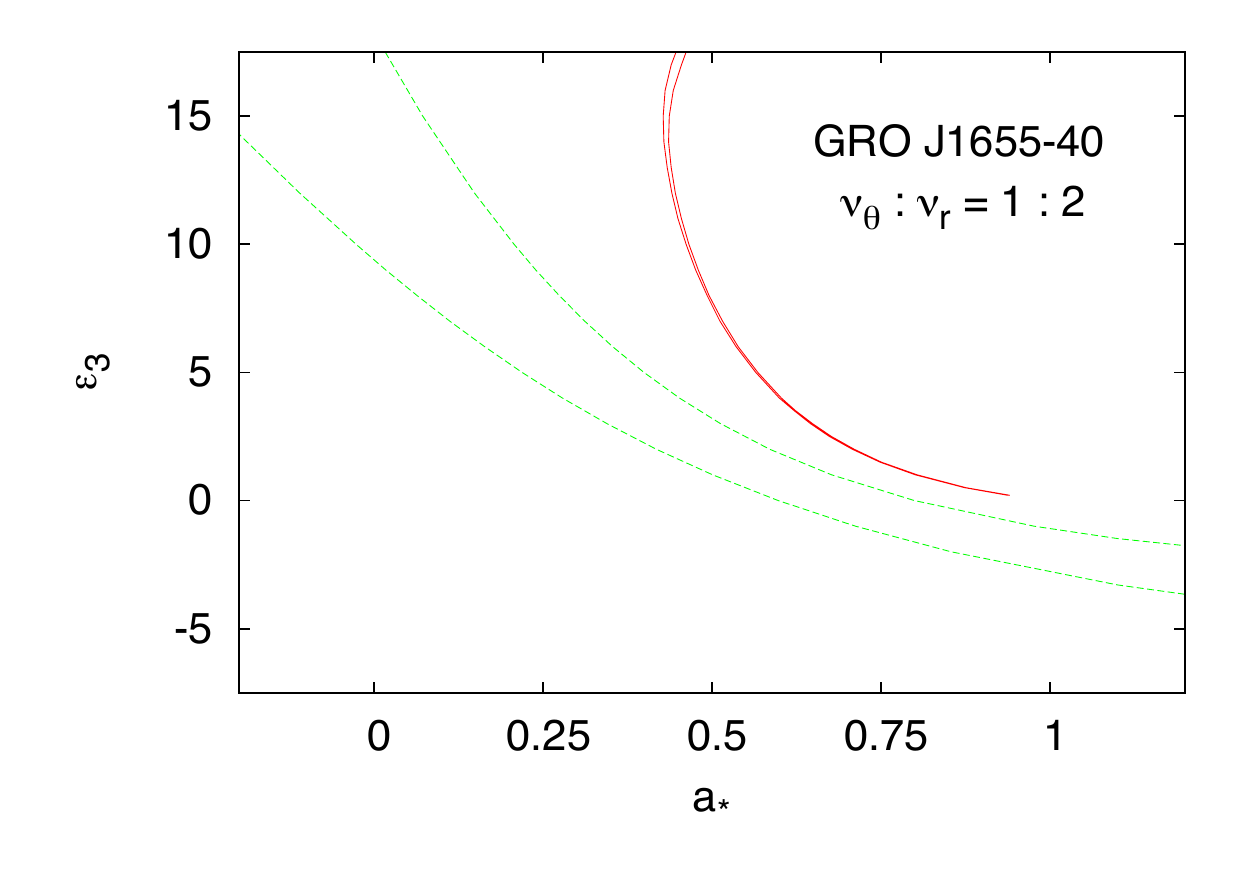}
\end{center}
\par
\vspace{-5mm} 
\caption{Constraints on the spin parameter $a_*$ and the JP deformation 
parameter $\epsilon_3$ for the BH candidate in the binary system GRO~J1655-40 
from resonance models of high-frequency QPOs (red solid curves) 
and the continuum-fitting method (green dashed curves). 
Top left panel: resonance $\nu_\theta : \nu_r = 3:2$ 
($\nu_{\rm U} = \nu_\theta$ and $\nu_{\rm L} = \nu_r$). 
Top right panel: resonance $\nu_\theta : \nu_r = 2:3$ 
($\nu_{\rm U} = \nu_r$ and $\nu_{\rm L} = \nu_\theta$).
Central left panel: resonance $\nu_\theta : \nu_r = 3:1$ 
($\nu_{\rm U} = \nu_\theta$ and $\nu_{\rm L} = \nu_\theta - \nu_r$).
Central right panel: resonance $\nu_\theta : \nu_r = 1:3$ 
($\nu_{\rm U} = \nu_r$ and $\nu_{\rm L} = \nu_r - \nu_\theta$).
Bottom left panel: resonance $\nu_\theta : \nu_r = 2:1$ 
($\nu_{\rm U} = \nu_\theta + \nu_r$ and $\nu_{\rm L} = \nu_\theta$).
Bottom right panel: resonance $\nu_\theta : \nu_r = 1:2$ 
($\nu_{\rm U} = \nu_r + \nu_\theta$ and $\nu_{\rm L} = \nu_r$).
The resonances $\nu_\theta : \nu_r = 2:3$, $1:3$, and $1:2$ (right panels) 
are not allowed in a Kerr space-time.}
\label{f-3}
\end{figure}

\begin{figure}
\par
\begin{center}
\includegraphics[type=pdf,ext=.pdf,read=.pdf,width=7.5cm]{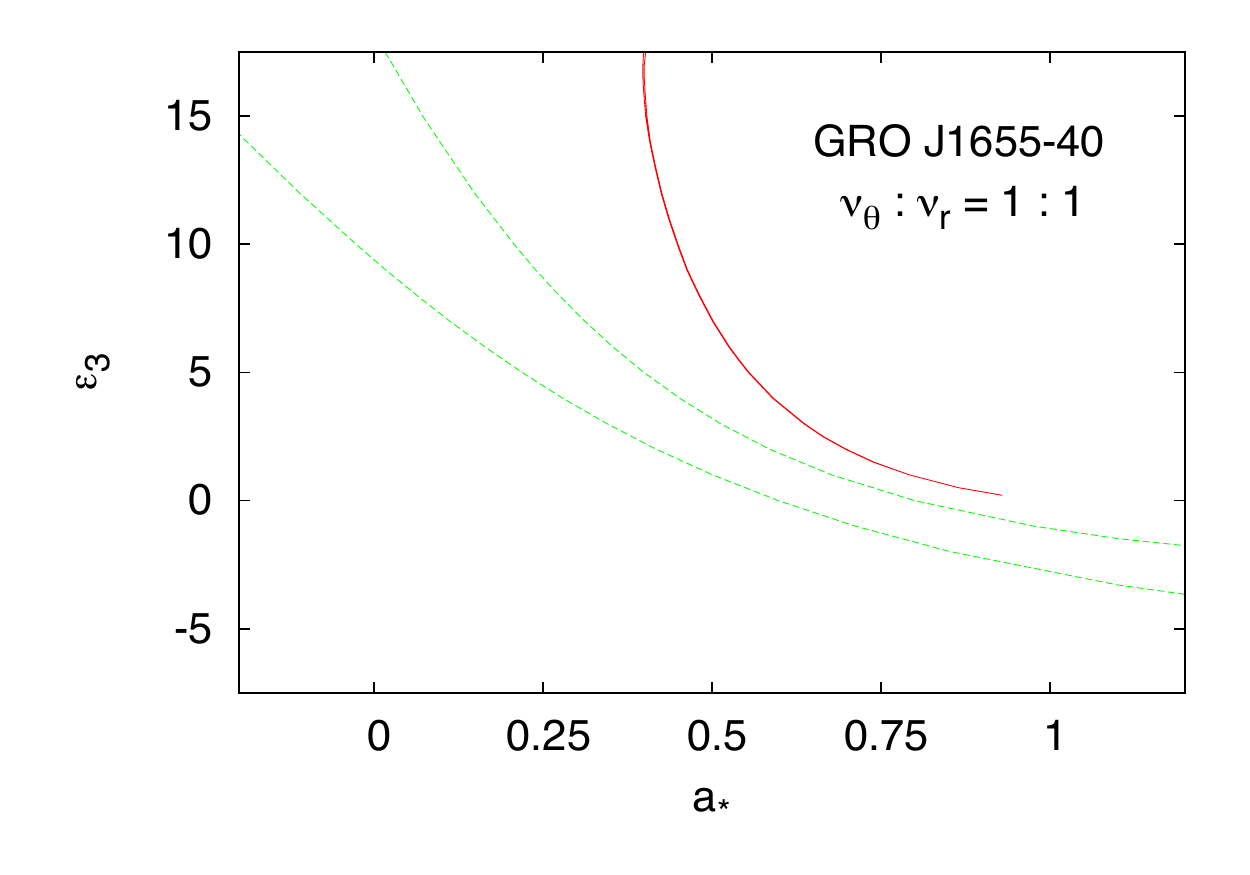}
\end{center}
\par
\vspace{-5mm} 
\caption{As in Fig.~\ref{f-3}, for the resonance $\nu_\theta : \nu_r = 1:1$, 
which is not allowed around a Kerr BH.}
\label{f-3a}
\end{figure}

\begin{figure}
\par
\begin{center}
\includegraphics[type=pdf,ext=.pdf,read=.pdf,width=7.5cm]{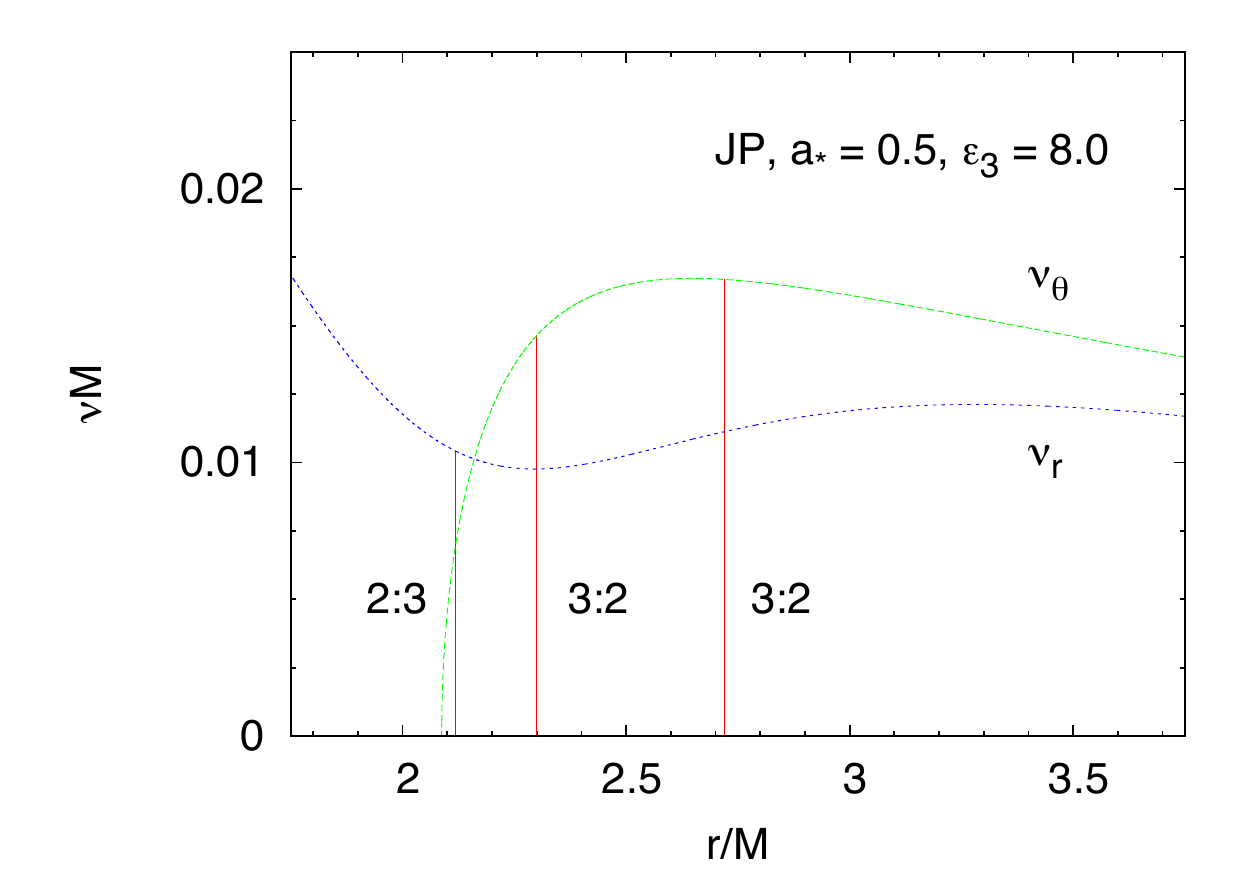}
\includegraphics[type=pdf,ext=.pdf,read=.pdf,width=7.5cm]{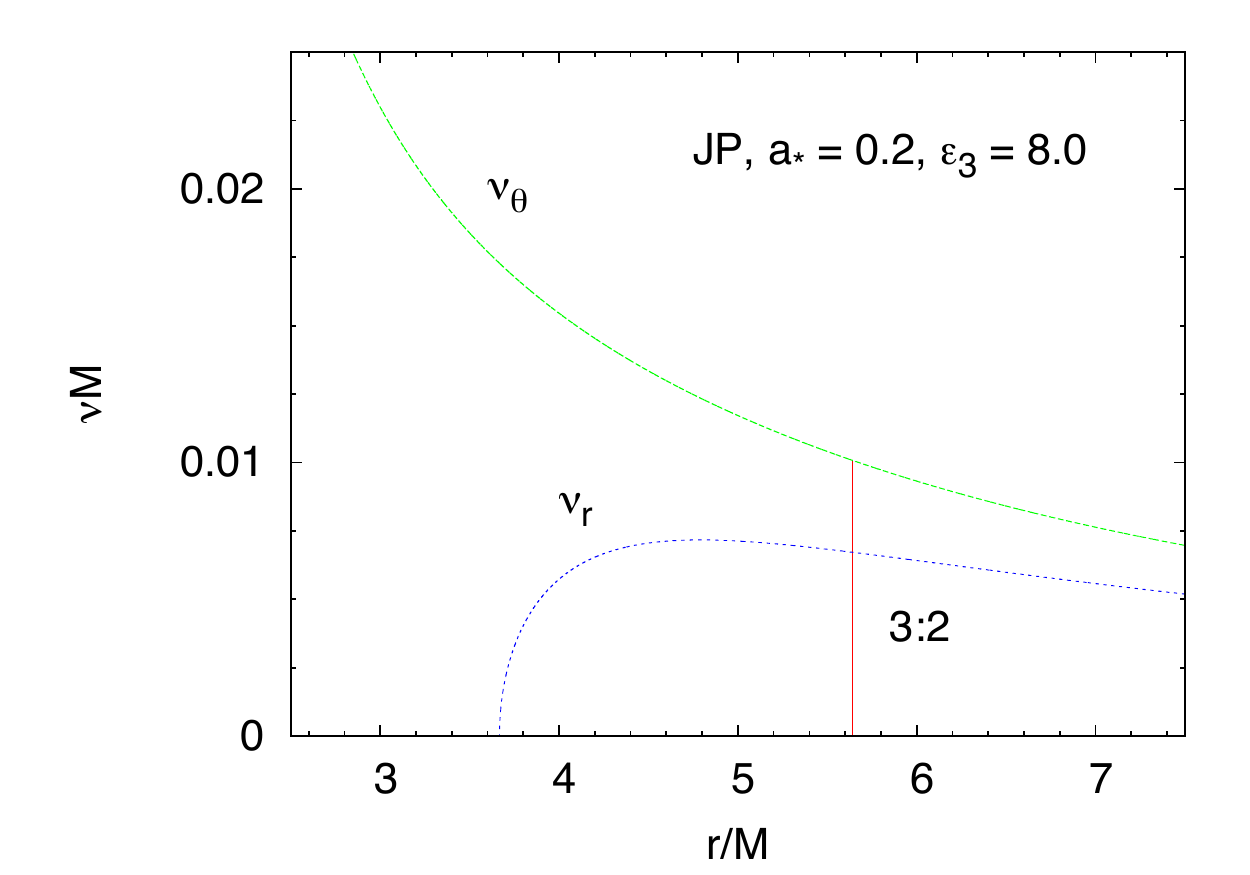} 
\end{center}
\par
\vspace{-5mm} 
\caption{Profiles of the radial epicyclic frequency $\nu_r$ and of the vertical 
epicyclic frequency $\nu_\theta$ in the JP space-time with deformation
parameter $\epsilon_3 = 8.0$. Left panel: spin parameter $a_* = 0.5$; 
the resonance $\nu_\theta : \nu_r = 3:2$ is present at the radii $r/M = 2.30$ 
and 2.72, while the resonance $\nu_\theta : \nu_r = 2:3$ is at $r/M = 2.12$. 
Right panel: spin parameter $a_* = 0.2$; there is only one resonance 
$\nu_\theta : \nu_r = 3:2$, which takes place at the radius $r/M = 5.64$, 
and there is no resonance $\nu_\theta : \nu_r = 2:3$.}
\label{f-3b}
\end{figure}

GRO~J1655-40 is the object with the best measurements for both the mass of the BH 
candidate and the two high-frequency QPOs, see Tab.~\ref{tab}. If we assume
$\nu_{\rm U}$ and $\nu_{\rm L}$ are the result of the coupling between the radial
and the vertical epicyclic oscillations, we can compute $\nu_r$ and $\nu_\theta$
from Eqs.~(\ref{eq-or}) and (\ref{eq-ot}), consider a particular choice of
$\{ m_1, m_2, n_1, n_2 \}$, and eventually find the allowed region on the plane
spin parameter-deformation parameter on the base of the observed $\nu_{\rm U}$ 
and $\nu_{\rm L}$. Considering only the uncertainty on the mass $M$ and
neglecting the one in $\nu_{\rm U}$ and $\nu_{\rm L}$, the allowed region for
some different choices of $\{ m_1, m_2, n_1, n_2 \}$ is shown in Figs.~\ref{f-3} and 
\ref{f-3a} (solid red curves)\footnote{When we constrain $a_*$ and $\epsilon_3$, 
we cannot restrict our study to the region $|a_*| \le 1$, as this bound is justified 
only for Kerr BHs~\cite{bh2,bh3}. In the case of non-Kerr objects, the issue of the 
instability depends on the details of the structure of the object and on the gravity 
theory, while the creation of a very compact object with $|a_*| > 1$ is possible
(in the specific case of the JP space-time, at least for $\epsilon_3 < 0$), as 
shown in~\cite{rad2,spin1,spin2}.}. Fig.~\ref{f-3b} shows the radial profiles of $\nu_r$ 
and $\nu_\theta$ for two JP space-times with $\epsilon_3 = 8.0$, one with 
$a_* = 0.5$ and the other one with $a_* = 0.2$. In the former case,
the condition $\nu_\theta : \nu_r = 3 : 2$ is satisfied at two radii and even the case
$\nu_\theta : \nu_r = 2 : 3$ is possible. In the second example, we have
$\nu_\theta : \nu_r = 3 : 2$ only at a single value of the radial coordinate and 
no $\nu_\theta : \nu_r = 2 : 3$.

\begin{figure}
\par
\begin{center}
\includegraphics[type=pdf,ext=.pdf,read=.pdf,width=7.5cm]{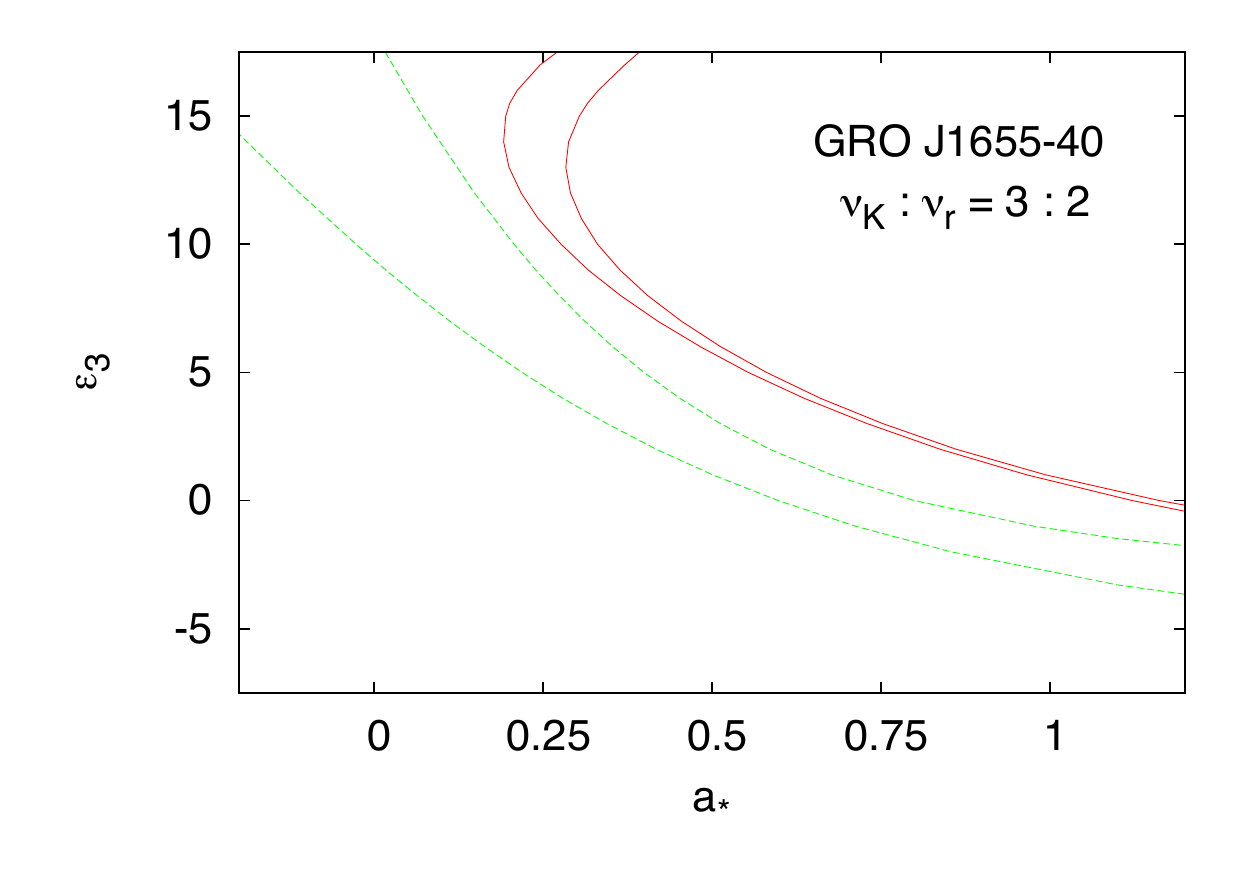} \\
\includegraphics[type=pdf,ext=.pdf,read=.pdf,width=7.5cm]{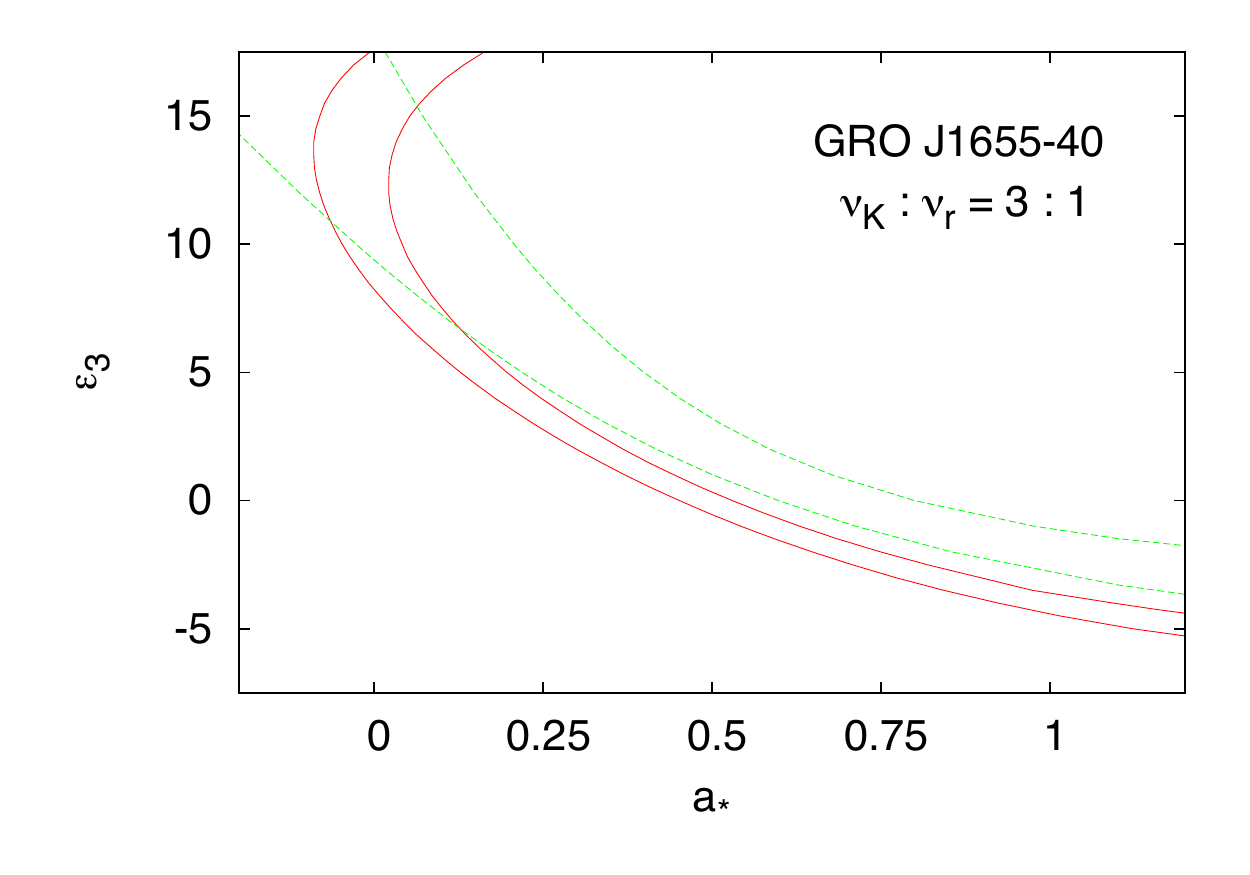}
\includegraphics[type=pdf,ext=.pdf,read=.pdf,width=7.5cm]{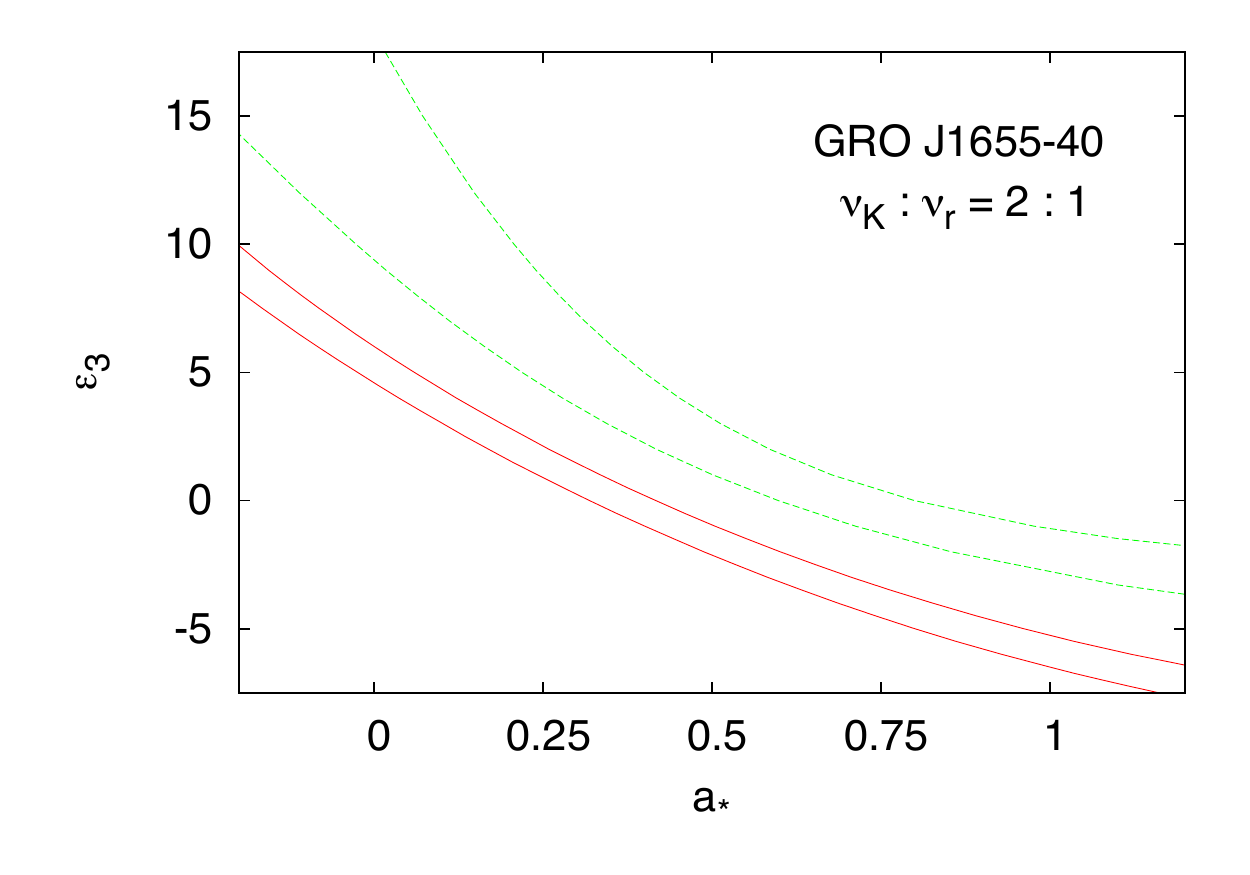} 
\end{center}
\par
\vspace{-5mm} 
\caption{As in Fig.~\ref{f-3}, in the case of Keplerian resonances: 
$\nu_{\rm K} : \nu_r = 3:1$ (top panel), $\nu_{\rm K} : \nu_r = 3:1$ (bottom left panel), 
and $\nu_{\rm K} : \nu_r = 2:1$ (bottom right panel).}
\label{f-4}
\end{figure}

As we can see, there is a degeneracy between the spin parameter $a_*$ and 
the deformation parameter $\epsilon_3$. This is the typical situation we find
when we relax the Kerr BH hypothesis and we allow for non-vanishing deviations
from the Kerr solution. In order to break this degeneracy, we need to use an
independent constraint on $a_*$ and $\epsilon_3$. For instance, we can use
the one coming from the analysis of the thermal spectrum of the accretion disk
of GRO~J1655-40 during the soft-high state. This technique is called continuum-fitting
method~\cite{cfmk1,cfmk2}, its validity is supported by a number of observational 
and theoretical studies~\cite{cfmk3}, and its extension to non-Kerr background is discussed 
in~\cite{cfm1,j1,j2}. In Figs.~\ref{f-3} and \ref{f-3a}, the region allowed by the 
continuum-fitting method is delimited by the dashed green line\footnote{The
constraints coming from the continuum-fitting method for GRO~J1655-40,
XTE~J1550-564, and GRS~1915+105 and shown in this paper are taken from 
Ref.~\cite{j1}.}.

We can repeat the same exercise to the cases with coupling between Keplerian 
and radial epicyclic frequencies. The simplest combinations are $\nu_{\rm K} : \nu_r
= 3:2$, $3:1$, and $2:1$. Fig.~\ref{f-4} shows the allowed regions in the
plane spin parameter-deformation parameter, with the green dashed line still
representing the bound coming from the continuum-fitting method.

\subsection{XTE~J1550-564}

\begin{figure}
\par
\begin{center}
\includegraphics[type=pdf,ext=.pdf,read=.pdf,width=7.5cm]{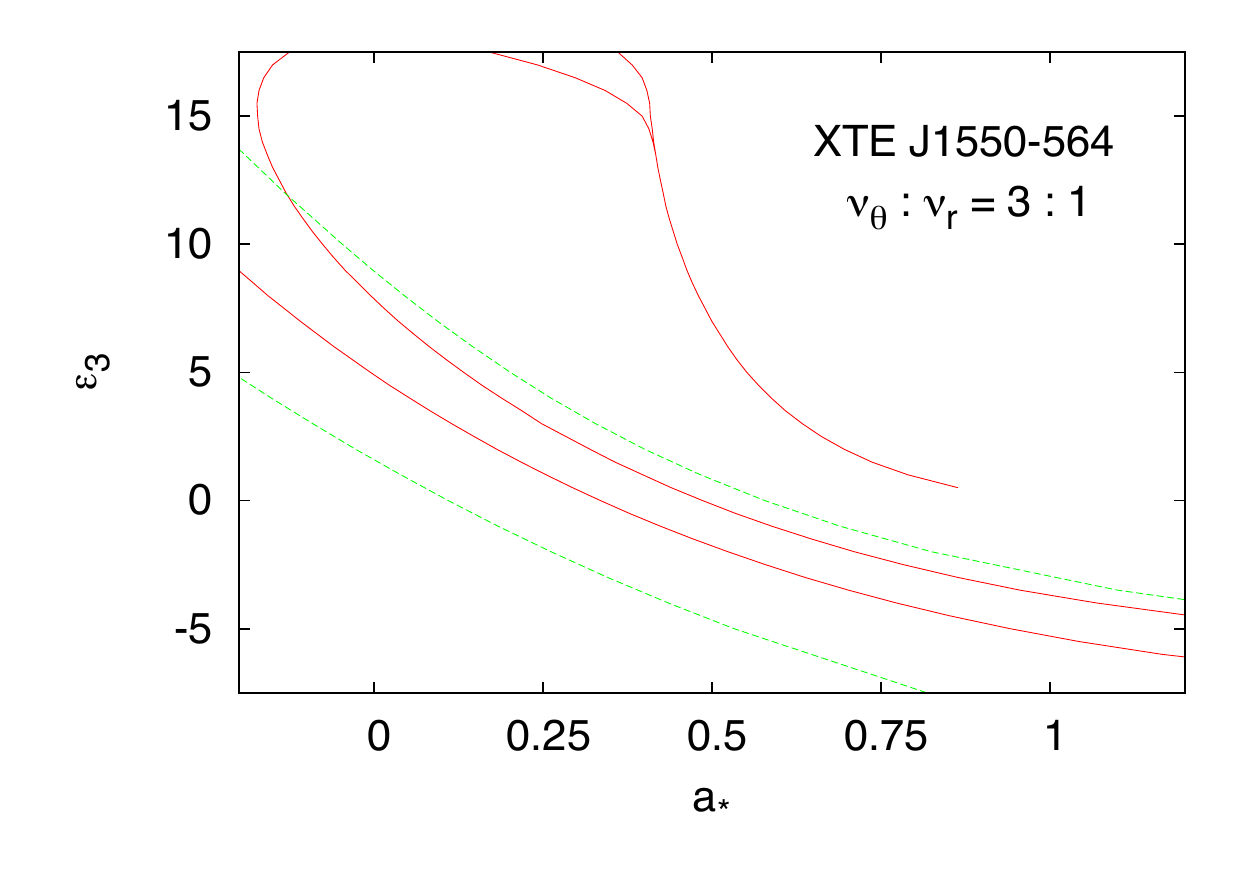}  
\includegraphics[type=pdf,ext=.pdf,read=.pdf,width=7.5cm]{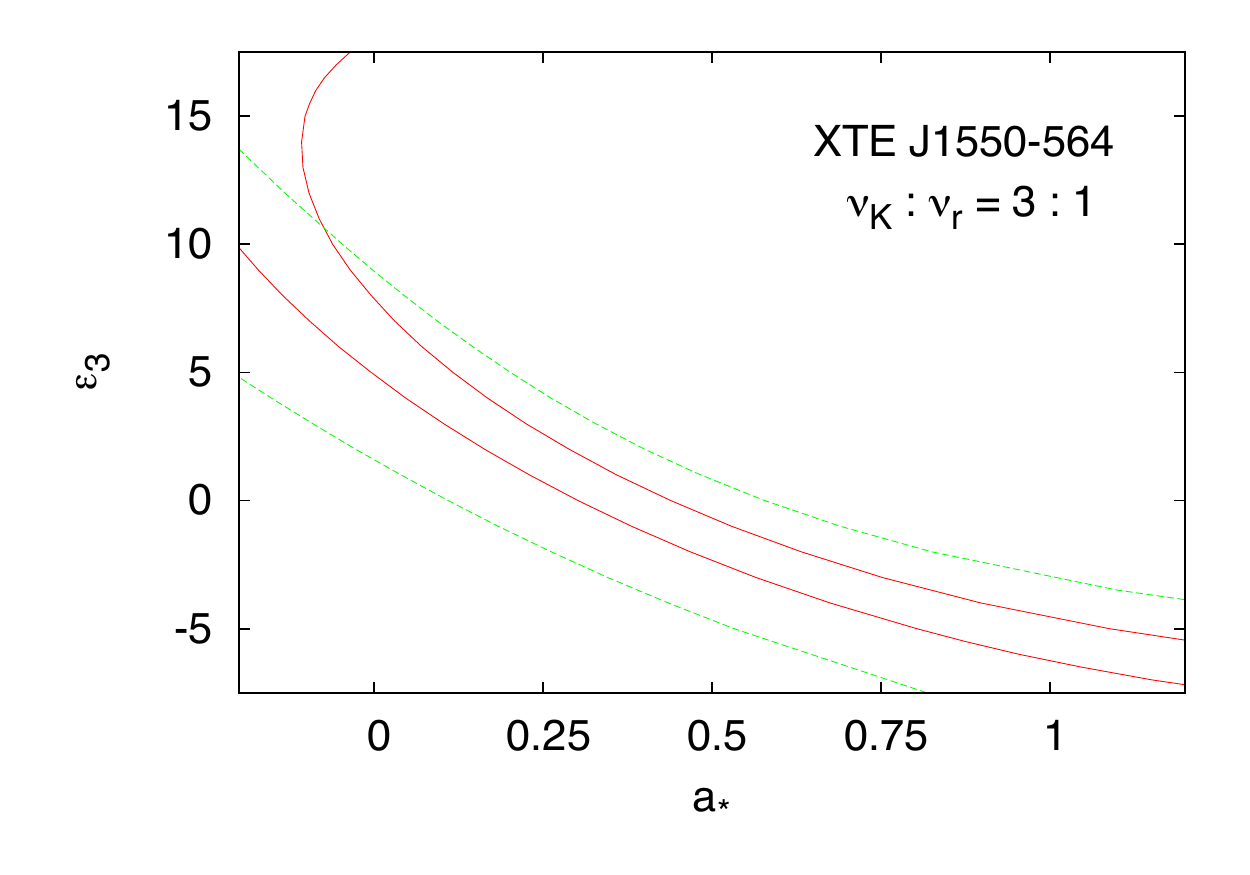}  
\end{center}
\par
\vspace{-5mm} 
\caption{Constraints on the spin parameter $a_*$ and the JP deformation 
parameter $\epsilon_3$ for the BH candidate in the binary system XTE~J1550-564
from resonance models of high-frequency QPOs (red solid curves) and the 
continuum-fitting method (green dashed curves). Resonances 
$\nu_\theta : \nu_r = 3:1$ (left panel) and $\nu_{\rm K} : \nu_r = 3:1$ (right panel).}
\label{f-5}
\end{figure}

For GRO~J1655-40, the resonance models considered in the previous subsection
and the continuum-fitting method seem to provide consistent results only for 
$\nu_\theta : \nu_r = 3:1$ and $\nu_{\rm K} : \nu_r = 3:1$ and for a non-vanishing 
deformation parameter $\epsilon_3 \sim 5 - 18$. For XTE~J1550-564, in addition to
these two possibilities, even the resonances $\nu_\theta : \nu_r = 2:1$ and 
$\nu_{\rm K} : \nu_r = 2:1$ would be in agreement with the results of the 
continuum-fitting method. However, in all the four cases the situation is like the
one shown in Fig.~\ref{f-5} and the degeneracy between $a_*$ and $\epsilon_3$ 
is not broken.

\subsection{GRS~1915+105}

\begin{figure}
\par
\begin{center}
\includegraphics[type=pdf,ext=.pdf,read=.pdf,width=7.5cm]{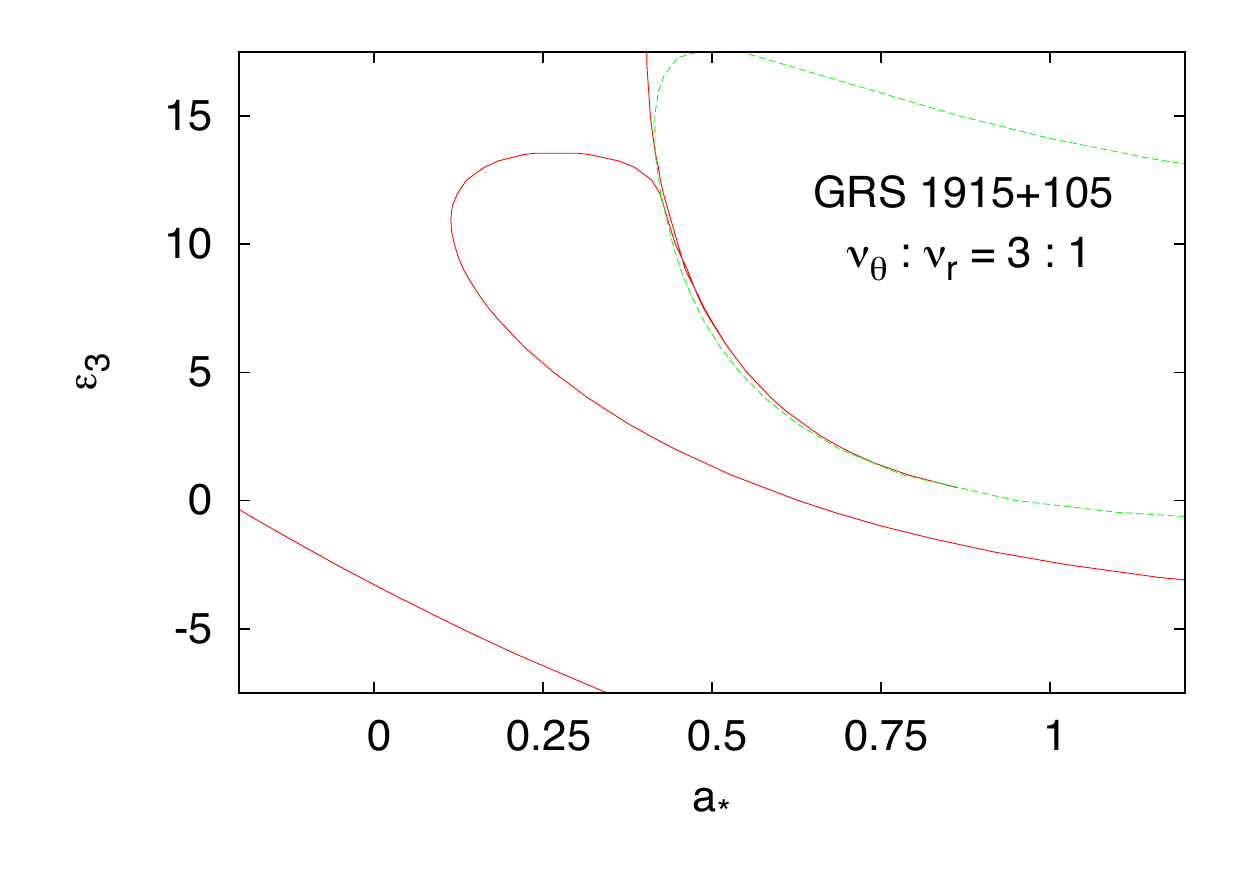}  
\includegraphics[type=pdf,ext=.pdf,read=.pdf,width=7.5cm]{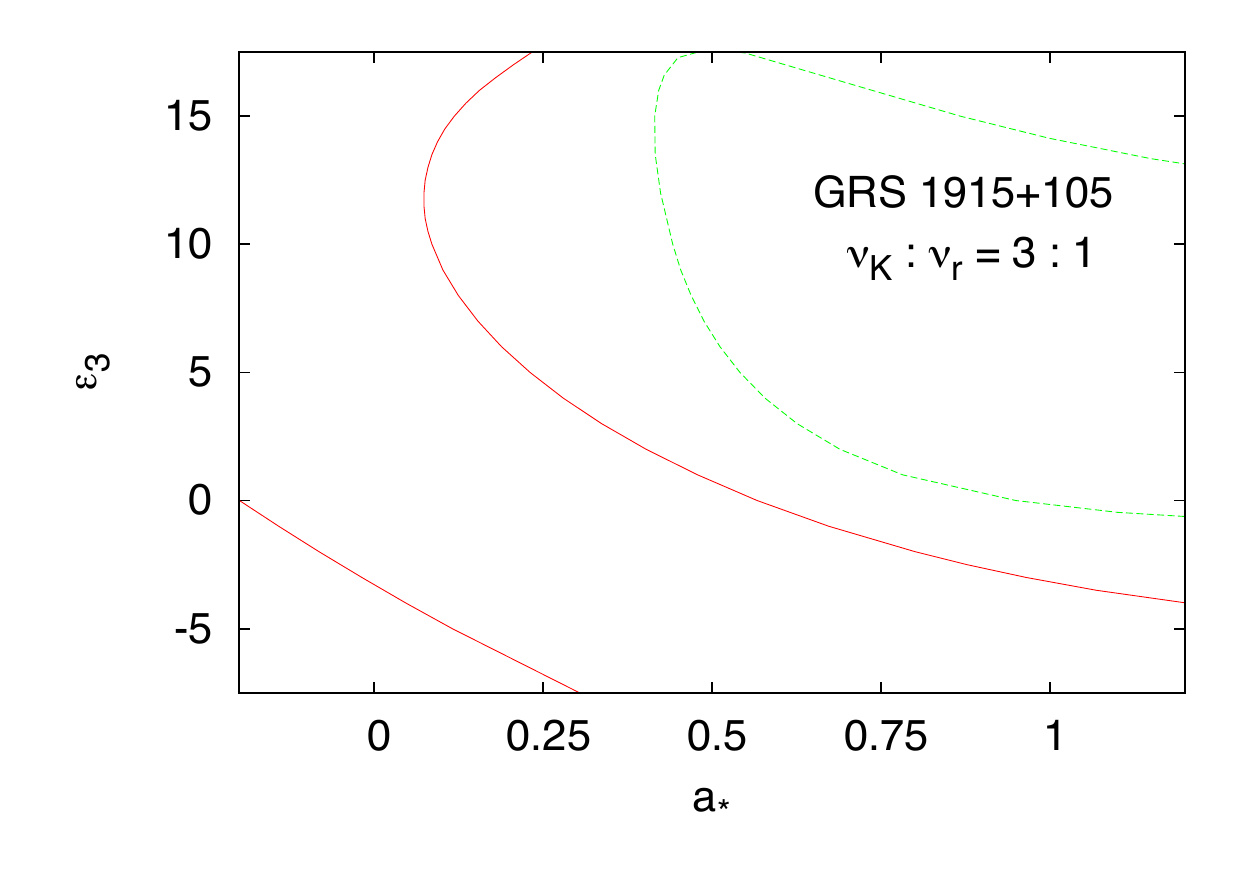}  
\end{center}
\par
\vspace{-5mm} 
\caption{Constraints on the spin parameter $a_*$ and the JP deformation parameter 
$\epsilon_3$ for the BH candidate in the binary system GRS~1915+105 from resonance 
models of high-frequency QPOs (red solid curves) and the continuum-fitting method 
(green dashed curves): resonances $\nu_\theta : \nu_r = 3:1$ (left panel) and 
$\nu_{\rm K} : \nu_r = 3:1$ (right panel).}
\label{f-6}
\end{figure}

For GRS~1915+105, the analysis of the thermal spectrum of the disk in the high-soft 
state in the Kerr background requires a quite high value of the spin parameter,
$a_* > 0.97$~\cite{nek}, which makes this objects very interesting\footnote{As for
GRO~J1655-40 and XTE~J1550-564, even in the case of GRS~1915+105 I am
considering the measurement obtained by the Harvard group. However, for 
GRS~1915+105 other groups have found different values, see e.g. Ref.~\cite{nek2}.}. 
If we consider the two resonances suggested by GRO~J1655-40 (i.e. $\nu_\theta : \nu_r 
= 3:1$ and $\nu_{\rm K} : \nu_r = 3:1$), we get the two panels in Fig.~\ref{f-6}. The 
resonance $\nu_\theta : \nu_r = 3:1$ (right panel in Fig.~\ref{f-6}) may still 
produce consistent results between the two approaches, while the resonance 
$\nu_{\rm K} : \nu_r = 3:1$ may be ruled out.

\begin{figure}
\par
\begin{center}
\includegraphics[type=pdf,ext=.pdf,read=.pdf,width=7.5cm]{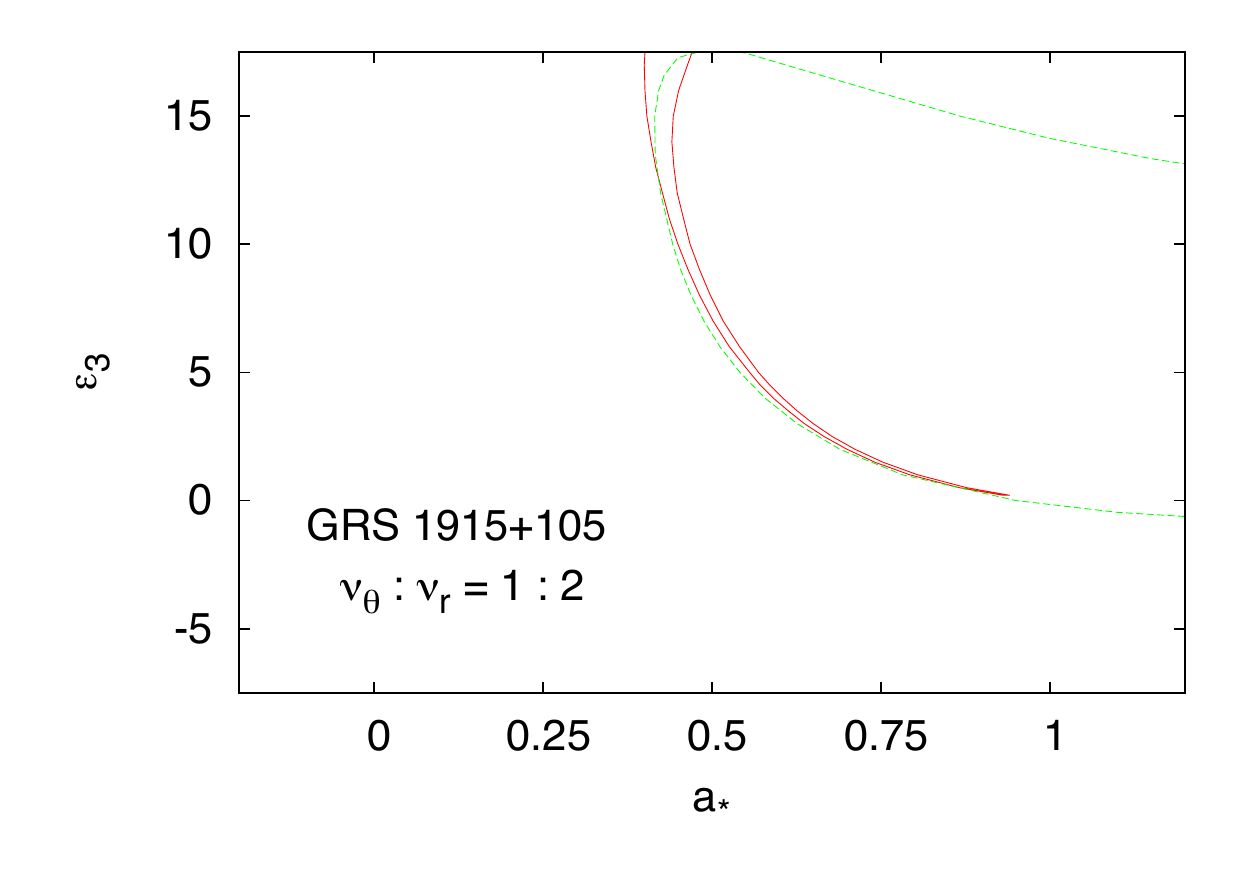}
\includegraphics[type=pdf,ext=.pdf,read=.pdf,width=7.5cm]{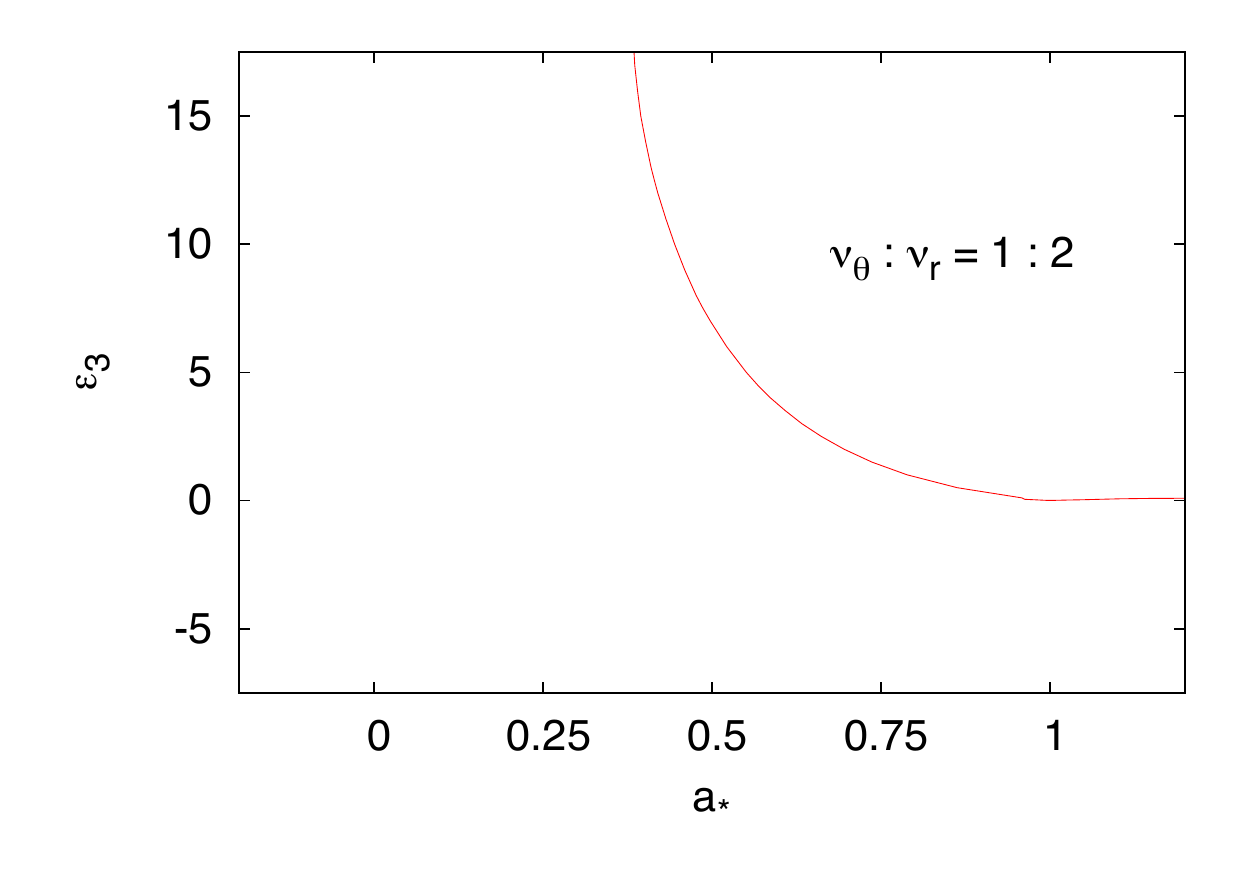}
\end{center}
\par
\vspace{-5mm} 
\caption{Left panel: as in Fig.~\ref{f-6}, for the resonance $\nu_\theta : \nu_r = 1:2$.
Right panel: boundary delimiting the JP space-times with and without the resonance
$\nu_\theta : \nu_r = 1:2$; for $\epsilon_3 \le 0$, the resonance $\nu_\theta : \nu_r = 1:2$
is never possible, while for $\epsilon_3 > 0$ the resonance exists only for some 
values of the spin parameter (we notice that the red curve reaches a minimum at 
$a_* = 1$, where $\epsilon_3 \rar 0^+$, and then goes up very slowly).}
\label{f-7}
\end{figure}

Fig.~\ref{f-7} shows instead the resonance $\nu_\theta : \nu_r = 1:2$, which is
allowed only for some values of $a_*$ and $\epsilon_3$ (see the right panel).
In this case, QPOs and continuum-fitting method provide consistent measurements
for non-vanishing $\epsilon_3$. That is true even for other resonances with
$\nu_r > \nu_\theta$, like $\nu_\theta : \nu_r = 1:3$ and $\nu_\theta : \nu_r = 1:1$ 
(not shown here). In a non-Kerr background we may thus have several
resonances that exist only if the compact object rotates sufficiently fast.

\section{Discussion and conclusions \label{s-5}}

The high-frequency QPOs observed in some stellar-mass BH candidates may
be used to test the Kerr-nature of these objects and to confirm, or rule out, the 
Kerr BH hypothesis. These frequencies are indeed almost constant, suggesting 
that their value is not determined by the
properties of the fluid accretion flow, but by the geometry of the space-time.
If we assume the stellar-mass BH candidates are the Kerr BHs predicted by 
General Relativity, the high-frequencies QPOs may be used to estimate the
spin parameter. However, at least in the framework of the
resonance models, the estimates of the spin parameter inferred from the observed 
QPOs of three micro-quasars seem not to be consistent with the measurements
obtained from the analysis of the thermal spectrum of the disk of the same 
objects, in the sense it is not possible to explain the observations of all the three 
micro-quasar by invoking a unique mechanism for the production
of the high-frequencies QPOs (actually, this assertion remains true even 
in all the other models proposed so far in the literature). 
We have thus four possibilities: $i)$ the resonance models are wrong,
$ii)$ the continuum-fitting method does not provide reliable estimate of $a_*$,
$iii)$ both techniques do not work correctly, $iv)$ these objects are not Kerr
BHs.

While systematic effects and/or wrong models are surely the most likely 
possibility to explain the disagreement between the two approaches, in this 
paper I explored the scenario $iv)$ and I found two (speculative) explanations:
\begin{enumerate}
\item {\it All the observations can be explained with the resonance 
$\nu_\theta : \nu_r = 3:1$}. As we can see from the central left panel in 
Fig.~\ref{f-3} and the left panels in Figs.~\ref{f-5} and \ref{f-6}, the resonance 
$\nu_\theta : \nu_r = 3:1$ may explain the high-frequency QPOs of the three 
micro-quasars in agreement with the measurements obtained from the 
continuum-fitting method, at the price of a deformation parameter $\epsilon_3 \sim 5-15$;
that is, these BH candidates should not be the Kerr BH predicted by General
Relativity. GRO~J1655-40 and XTE~J1550-564 would be slow-rotating compact
objects and the profile of the radial and the vertical epicyclic frequencies 
would be like the one in the right panel of Fig.~\ref{f-3b}; that is, the resonance
$\nu_\theta : \nu_r = 3:1$ would be present at a unique radius $r_{3:1}$.
GRS~1915+105 would instead be a ``fast-rotating'' object with $a_* \approx 0.5$
and the situation would be like the one in the left panel of Fig.~\ref{f-3b}:
the resonance $\nu_\theta : \nu_r = 3:1$ is possible at two different radii,
but we would observe the one occurring at the smaller one, which should
indeed be expected to be stronger. Let us notice that the value $a_* \approx 0.5$
might be close to the maximum value of the spin parameter for a compact 
object with $\epsilon_3 \sim 5-15$, see Fig.~1 in Ref.~\cite{rad2}. For
GRS~1915+105, the allowed region delimited by the green dashed line is
likely in large part unphysical (equivalent to a Kerr metric with $|a_*| > 1$, i.e.
it may be impossible to create such a fast-rotating objects) and therefore it is not so strange
that the overlap between the bounds from the QPOs and the continuum-fitting
method is a thin area near the boundary of the region allowed by the 
continuum-fitting method. 
\item {\it The observed high-frequency QPOs are produced by different resonances
determined by value of the the spin parameter of the compact object}. 
Unlike in the Kerr background, for non-Kerr
metrics such a scenario is much more appealing because there are excitation 
modes possible only if the object has a spin parameter exceeding a critical
value. So, for a given deformation parameter, the resonance responsible for
the production of the high-frequency QPOs would be determined by the specific
value of the spin parameter of the compact object. This possibility can be better
understood with the following example. In the model briefly reviewed in 
Subsection~\ref{ss-me}, resonances are possible only for $\nu_r/\nu_\theta = 2/n$ 
with $n$ integer. When $\nu_\theta > \nu_r$, the minimum $n$ is 3. However, 
for $\epsilon_3 > 0$, the minimum value of $n$ depends on the spin parameter 
$a_*$. For small value of the spin parameter, $n_{\rm min} = 3$, as in Kerr. Above 
a critical value of the spin parameter, which depends on $\epsilon_3$, 
$n_{\rm min} = 1$ or 2. So, we may imagine a model in which GRO~J1655-40 
and XTE~J1550-564 present the resonance $\nu_\theta : \nu_r = 3:1$, because 
their BH candidates are slow-rotating objects and $\nu_\theta > \nu_r$. On the 
other hand, for GRS~1915+105 even the resonances with $\nu_r > \nu_\theta$
could be excited, and actually be stronger. A deformation parameter 
$\epsilon_3 \sim 5-15$ is still required, i.e. the three 
micro-quasars should be objects more prolate than Kerr BHs.
\end{enumerate}

The high-frequency QPOs are potentially an excellent tool to investigate the
geometry of the space-time around stellar-mass BH candidates and to measure
the fundamental properties of these objects. For the time being, however, we do not
know the exact origin of these QPOs and therefore any analysis depends on the
assumed model. In this paper, I have considered the resonance models first
proposed by Abramowicz and Kluzniak, because they have a number of nice
features, and compared the results with the measurements of the thermal
spectrum of the disk performed by the Harvard group. All the observations of the
three micro-quasars for which we can do this test can be explained if we admit
that these objects are not Kerr BHs. Within the theoretical framework proposed
by Johannsen and Psaltis with a single deformation parameter, one finds
that current data demand $\epsilon_3 \sim 5-15$, while a Kerr BHs would
require $\epsilon_3 = 0$, and for $\epsilon_3 > 0$ ($< 0$) the gravitational force
on the equatorial plane is weaker (stronger) than the one around a Kerr BH
with the same mass and spin. At a speculative level, we can notice that this
value of $\epsilon_3$ is consistent with another non-zero measurement,
$\epsilon_3 \approx 7.5$, found in Ref.~\cite{j2}. The idea behind the work
of Ref.~\cite{j2} is that the steady jets observed in some micro-quasars
may be powered by the Blandford-Znajek mechanism~\cite{znajek}, which is 
one of the most appealing scenarios to explain this kind of phenomena.
If this is the case, one should expect a correlation between the
power of steady jets and the value of the spin parameter of
the micro-quasars. Comparing current estimates of the jet power with
the measurements of the spin parameter obtained by the continuum-fitting
method, one finds there is no correlation if the micro-quasars are Kerr BHs,
i.e. if $\epsilon_3 = 0$, but a correlation is possible, and actually close to
the theoretical predictions $P_{\rm jet} \sim a_*^2$, for $\epsilon_3 \approx 7.5$~\cite{j2}.


\acknowledgments
This work was supported by the Humboldt Foundation.


\end{document}